\documentclass[aps,pra,twocolumn,amsmath,amssymb,superscriptaddress,longbibliography]{revtex4-2}
\usepackage{bm,psfrag,setspace,braket}
\usepackage{hhline}
\usepackage[T1]{fontenc}
\usepackage[left]{lineno}
\usepackage[pdftex]{graphicx} 
\selectfont

\newcommand{\bracenom}{\genfrac{\lbrace}{\rbrace}{0pt}{}}

\begin{document}
\title{Exact $k$-body representation of the Jaynes-Cummings interaction in the dressed basis: Insight into many-body phenomena with light}
\author{Kevin C. Smith}
\affiliation{Department of Physics, University of Washington, Seattle, Washington 98195-1560, USA}
\author{Aniruddha Bhattacharya}
\affiliation{Department of Chemistry, University of Washington, Seattle, Washington 98195-1700, USA}
\author{David J. Masiello}
\email{masiello@uw.edu}
\affiliation{Department of Chemistry, University of Washington, Seattle, Washington 98195-1700, USA}

\begin{abstract}
Analog quantum simulation -- the technique of using one experimentally well-controlled physical system to mimic the behavior of another -- has quickly emerged as one of the most promising near term strategies for studying strongly correlated quantum many-body systems. In particular, systems of interacting photons, realizable in solid-state cavity and circuit QED frameworks, for example, hold tremendous promise for the study of nonequilibrium many-body phenomena in part due to the capability to locally create and destroy photons. These systems are typically modeled using a Jaynes-Cummings-Hubbard (JCH) Hamiltonian, named due to similarities with the Bose-Hubbard Hamiltonian. While comparisons between the two are often made in the literature, the JCH Hamiltonian comprises both bosonic and psuedo-spin operators, leading to physical deviations from the Bose-Hubbard model for particular parameter regimes. Here, we present a non-perturbative procedure for transforming the Jaynes-Cummings Hamiltonian into a dressed operator representation that, in its most general form, admits an infinite sum of bosonic $k$-body terms where $k$ is bound only by the number of excitations in the system. We closely examine this result in both the dispersive and resonant coupling regimes, finding rapid convergence in the former and contributions from $k\gg1$ in the latter. Through extension to the simple case of a two-site JCH system, we demonstrate that this approach facilitates close inspection of the analogy between the JCH and Bose-Hubbard models and its breakdown for resonant light-matter coupling. Finally, we use this framework to survey the many-body character of a two-site JCH for general system parameters, identifying four unique quantum phases and the parameter regimes in which they are realized,  thus highlighting phenomena realizable with finite JCH-based quantum simulators beyond the Bose-Hubbard model. More broadly, this work is intended to serve as a clear mathematical exposition of bosonic many-body interactions underlying Jaynes-Cummings-type systems, often postulated either through analogy to Kerr-like nonlinear susceptibilities or by matching coefficients to obtain the appropriate eigenvalue spectrum.
\end{abstract}
\maketitle

\section{\label{sec:intro}Introduction}
Efficient simulation of strongly correlated many-body systems remains one of the most important unsolved problems in the physical sciences today, promising advances in a diverse set of fields ranging from high-energy physics and cosmology to quantum chemistry and condensed matter physics \cite{preskill2018quantum,altman2021quantum}. It is also one of the most challenging, as such systems involve dynamics within a Hilbert space whose size increases exponentially with added degrees of freedom, rendering brute force study of many-body systems impractical with even the most powerful classical computers. Feynman famously recognized this problem nearly four decades ago and proposed what is now termed a quantum simulator -- a programmable machine whose underlying degrees of freedom are quantum mechanical, circumventing the exponential scaling problem and thus enabling efficient simulation of quantum systems \cite{feynman1982simulating,10.2307/2899535,abrams1997simulation,abrams1999quantum}. These devices generally fall into two classes: digital and analog quantum simulators. The former are an application of universal quantum computers which, despite rapid advancement in recent years, are likely decades away from a practical, fault-tolerant realization \cite{hauke2012can,cirac2012goals,mcclean2020decoding,preskill2018quantum,campbell2017roads}. In contrast, the latter are specialized, comparatively less ambitious devices comprising a well-controlled quantum system which mimics a particular quantum system of interest with some degree of tunability \cite{altman2021quantum, kokail2019self}. Analog quantum simulators thus offer a viable near-term solution for study of quantum many-body phenomena, and consequently a wide array of physical systems have been experimentally and theoretically studied as platforms for analog quantum simulation in recent years \cite{georgescu2014quantum, buluta2009quantum, jaksch2005cold, lewenstein2007ultracold, bloch2012quantum, blatt2012quantum, schneider2012experimental, aspuru2012photonic, houck2012chip,cirac2012goals,altman2021quantum,lamata2018digital,braumuller2017analog,hensgens2017quantum,lv2018quantum,arguello2019analogue}.

One of the most unique classes of proposed platforms entails emulation of quantum many-body physics with light. As photons do not naturally interact, replicating an interacting many-body system relies on experimental realization of single-photon nonlinearities, a difficult task particularly in the optical domain. In cavity and circuit QED settings, one strategy for achieving nonlinearity involves realization of the Jaynes-Cummings model, which describes a single quantized cavity mode interacting with a two-level system (TLS). If the rate of dissipation to the environment is exceeded by the rate of coherent energy exchange between the cavity mode and TLS, the system is said to be in the strong coupling regime and a phenomenon known as photon blockade can occur whereby absorption of a single photon of a particular frequency prevents further absorption at that same frequency, thus enabling single photon nonlinearity and, consequently, Kerr-type photon-photon interactions \cite{imamoglu1997strongly,grangier1998comment,birnbaum2005photon,faraon2008coherent,lang2011observation,hoffman2011dispersive}. A suitable platform for quantum simulation is then realized by an array of TLS-enabled nonlinear cavities, where the pure photonic modes of adjacent cavities are coupled through the mutual overlap of their evanescent fields. Such a system shares similarities with the Bose-Hubbard model and is commonly referred to as the Jaynes-Cummings-Hubbard (JCH) model \cite{greentree2006quantum,hartmann2006strongly,Angelakis2007, koch2009superfluid, grujic2012non}, combining Hubbard-like on-site interactions (mediated by the TLS) with bosonic hopping between adjacent sites. 

Unlike other notable quantum simulation platforms, such as those composed of ultracold atoms in optical lattices \cite{greiner2002quantum,gross2017quantum,tarruell2018quantum}, an array of TLS-enabled nonlinear cavities does not provide an exact analog of the Bose-Hubbard model. For one, the JCH Hamiltonian is composed of both bosonic and psuedospin operators, while the Bose-Hubbard Hamiltonian contains only the former. In addition, whereas the insulator-to-superfluid phase transition of the Bose-Hubbard model is understood through analysis of the competition between on-site repulsion $U$ and hopping strength $J$, the various phases of the JCH model are determined by three competing energy scales: on-site repulsion $U$, hopping strength $J$, and TLS-cavity detuning $\Delta$. Despite these differences, it has been shown that the JCH model admits an insulator-to-superfluid phase transition much like that of the Bose-Hubbard model \cite{greentree2006quantum,hartmann2006strongly,hartmann2008quantum,koch2009superfluid,Angelakis2007,Noh2016,PhysRevLett.99.186401,PhysRevA.77.033801,PhysRevA.77.053819} and, consequently, the two have been closely compared in a number of publications \cite{greentree2006quantum,hartmann2006strongly,hartmann2007strong,hartmann2008quantum,leib2010bose,koch2009superfluid,schmidt2009strong,Angelakis2007,carusotto2009fermionized,hohenadler2011dynamical,PhysRevLett.99.186401,PhysRevA.77.033801,PhysRevA.77.053819,hartmann2016quantum,PhysRevLett.111.160501,mering2009analytic,nietner2012ginzburg,bujnowski2014supersolid,hayward2012fractional}.

Here, we present a thorough analysis of the many-body character underlying the Jaynes-Cummings Hamiltonian and ultimately revisit the analogy between the JCH and Bose-Hubbard models for the simplest possible implementation: a two-site system. We begin by considering just a single Jaynes-Cummings system and introduce a parameter-independent strategy for exposing an infinite hierarchy of bosonic many-body interactions at the level of dressed operators. In contrast to similar methods prominent in the literature \cite{boissonneault2009dispersive, blais2004cavity, blais2020circuit}, our approach is non-perturbative and is therefore valid for general system parameters, facilitating analysis of both dispersive and resonant light-matter coupling regimes and providing explicit mathematical relations between the parameters and operators appearing in the Jaynes-Cummings Hamiltonian and its many-body representation. We apply this methodology toward analysis of a two-site JCH model in both photonic and polaritonic regimes, and show that the dispersive coupling limit provides an approximate analog to the Bose-Hubbard model, albeit with weak nonlinearities. In the resonant regime, we illustrate the disconnect between these two models, and make explicit the polaritonic-to-photonic transition which accompanies the two-site analog to the familiar insulator-to-superfluid transition of the Bose-Hubbard model. To that end, we show that the resonant coupling case admits of a third, intermediary phase consistent with a polaritonic superfluid, highlighting the distinct possibilities afforded by the JCH model over the Bose-Hubbard case. Taken together, the goal of this paper is to present a unique, parameter-independent approach for studying the effective many-body interactions realizable in Jaynes-Cummings-type systems and, via extension to a two-site system, fully explore the various parameter regimes of a simple, finite Jaynes-Cummings-Hubbard system with an eye towards experimental study of many-body phenomena using photonics-based platforms.

The subsequent sections are organized as follows. In Section \ref{sec:JC} we derive a many-body representation for the Jaynes-Cummings Hamiltonian in terms of dressed operators and discuss its limiting cases for various parameter regimes. This is carried out in three parts: Section \ref{subsec:JC_diag} contains a derivation of the dressed operator representation of the Jaynes-Cummings Hamiltonian, followed by a discussion of the behavior of the dressed operators in \ref{subsec:dressedops} and, in Section \ref{subsec:manybodyrep}, a derivation and analysis of our main result -- an exact, many-body representation of the Jaynes-Cummings Hamiltonian. Section \ref{sec:2JCH} extends our methods to the two-site JCH model, beginning with a brief comparison between the Bose-Hubbard and JCH models in Section \ref{subsec:JCH_BH_comp}. This is followed by a more thorough analysis of the two-site JCH in the dispersive and two excitation limits in Sections \ref{subsec:2JCHdisp} and \ref{subsec:2JCH_neq2}. We then examine the various quantum phases of the two-site JCH in Section \ref{subsec:2JCHphases} before concluding with a summary of our findings in Section \ref{sec:Concl}.

\section{\label{sec:JC}Non-perturbative Many-Body Representation of The Jaynes-Cummings Hamiltonian}

\begin{figure}
\includegraphics[width=0.4\textwidth]{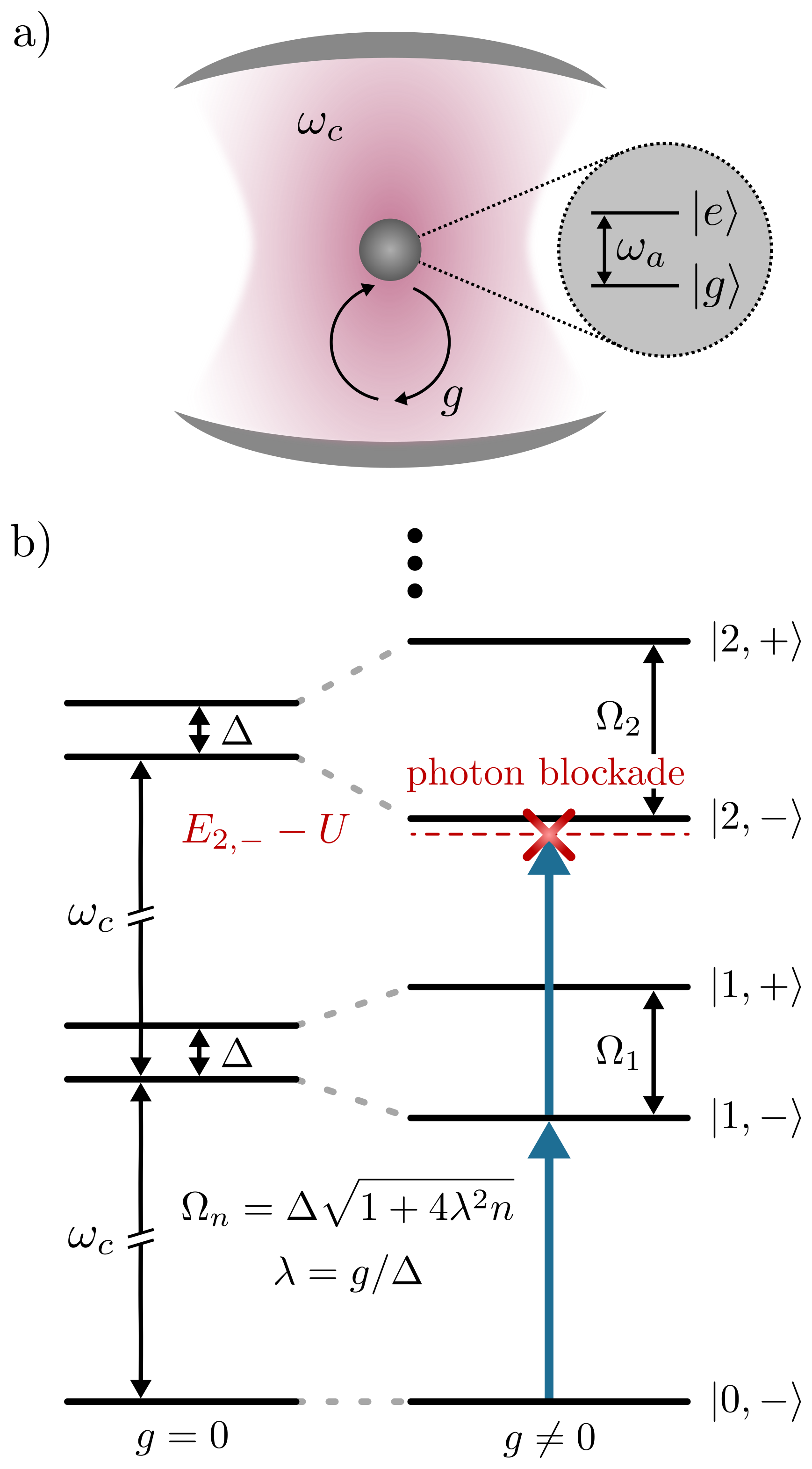}
\caption{\label{fig:f1} (a) A realization of the Jaynes-Cummings model consisting of a single cavity mode and a two-level system (TLS) of resonant frequency $\omega_c$ and $\omega_a$, respectively. The two interact, exchanging quanta at a rate determined by the coupling strength $g$. (b) The eigenspectrum of the Jaynes-Cummings Hamiltonian for $\Delta>0$. The left set of horizontal black lines indicates the eigenenergies of the bare cavity mode and TLS, while the right portrays the impact of light-matter coupling. Pairs of states with the same total number of excitations $n$ hybridize, yielding pairs of dressed eigenstates $\ket{n,\pm}$ which are split by the frequency $\Omega_n$. If the system begins in the ground state $\ket{0,-}$, absorption of one photon of frequency $(E_{1,-}-E_{0,-})$ prohibits absorption of a second of the same frequency due to the additional energy cost $U$. This phenomenon is known as photon blockade, and may be used to realize effective photon-photon interactions.}
\end{figure}

We begin by examining the hidden bosonic many-body nature of the Jaynes-Cummings Hamiltonian, one of the simplest and most versatile models in quantum optics describing the coherent interaction between a single cavity mode and a TLS, as shown in Fig. \ref{fig:f1}. Defining $a^\dagger$ and $a$ as creation and annihilation operators for the bosonic cavity mode and $\sigma^+=\ket{e}\bra{g}$ and $\sigma^-=\ket{g}\bra{e}$ as psuedo-spin raising and lowering operators describing transitions between the ground $\ket{g}$ and excited $\ket{e}$ states of the TLS, the Jaynes-Cummings Hamiltonian is given by
\begin{equation}
    H = \hbar\omega_c a^\dagger a + \frac{1}{2}\hbar\omega_a\sigma^z + \hbar g (a^\dagger \sigma^- + a \sigma^+).
\label{eq:JC}
\end{equation}
Here, $\omega_c$ is the resonant frequency of the cavity mode and $\omega_a$ that of the TLS or ``atom'' - terminology which will be used interchangeably for the remainder of this work. We emphasize that the physical implementation of the TLS need not be an atom, and may instead describe the energy levels of a so-called \emph{artificial} atom such as a superconducting qubit \cite{wallraff2004strong, Schoelkopf2008, Haroche2020, blais2020circuit} or quantum dot \cite{yoshie2004vacuum,reithmaier2004strong,hennessy2007quantum, haroche2006exploring}. The rate of energy exchange between the cavity and TLS is defined by the coupling strength $g$, here assumed to be fast enough such that the atom and cavity are strongly coupled and dissipation may be neglected at first approximation \cite{cohen1998atom, haroche2006exploring}, yet not so fast that the counter-rotating terms of the Rabi model be considered (i.e., $g\ll\{\omega_c,\omega_a\}$) \cite{Kockum2019,casanova2010deep}. Finally, $\sigma^z$ is the Pauli operator $\sigma^z = [\sigma^+,\sigma^-] = \ket{e} \bra{e}-\ket{g}\bra{g}$.

The eigenvectors and eigenvalues of Eq. (\ref{eq:JC}) are most easily found by recognizing that the Hamiltonian conserves the total number of quanta 
\begin{equation}
    N=a^\dagger a + \sigma^+\sigma^-.
\end{equation} Consequently, only states within the same excitation number manifold couple and Eq. (\ref{eq:JC}) may be rewritten as a block-diagonal matrix with each $2\times2$ block independently diagonalizable. Following Ref.~\onlinecite{boissonneault2009dispersive}, we define
\begin{equation}
    \lambda = g/\Delta
\end{equation}
where $\Delta=\omega_a-\omega_c$ is the atom-cavity detuning. Then the eigenvalues may be written as
\begin{equation}
\begin{split}
    E_{n,\pm} &= \left(n-\frac{1}{2}\right)\hbar\omega_c \pm \frac{\hbar}{2}\Delta\sqrt{1+4 \lambda^2n}\\
    \end{split}
\label{eq:JCeigenval}
\end{equation}
with associated eigenvectors
\begin{equation}
    \begin{split}
        \ket{n,-} &= \cos\theta\ket{n,g}-\sin\theta \ket{n-1,e} \\
        \ket{n,+} &= \sin\theta\ket{n,g}+\cos\theta \ket{n-1,e},
    \end{split}
\label{eq:JCeigenvec}
\end{equation}
where $n$ are eigenvalues of the total number operator $N$ which label the excitation manifold and the mixing angle $\theta=\tan^{-1}(2\lambda\sqrt{n})/2$
describes the degree of hybridization between photonic and atomic degrees of freedom, taking values in the range $-\pi/4\leq\theta\leq\pi/4$ with upper and lower bounds corresponding to maximal mixing and $\theta=0$ indicating an uncoupled system.
 
 The eigenspectrum of Eq. (\ref{eq:JC}) is often referred to as the Jaynes-Cummings ladder \cite{fink2008climbing,kasprzak2010up,laussy2012climbing,hopfmann2017transition}, shown in Fig. \ref{fig:f1}b. Crucially, this spectrum is nonlinear in $n$, leading to a phenomenon known as \emph{photon blockade} \cite{imamoglu1997strongly, birnbaum2005photon}, whereby absorption of a photon at a particular frequency inhibits further absorption of photons at that same frequency (see Fig. \ref{fig:f1}c). In this way, the Jaynes-Cummings Hamiltonian facilitates \emph{effective} photon-photon interactions in the few photon limit. Caution must be exercised, however, in attempting to write down an effective Hamiltonian which accounts for these effects. In particular, it is clear from Fig. \ref{fig:f1}b that transition to the state $\ket{2,-}$ through absorption of successive photons of frequency $\omega = (E_{1,-}-E_{0,-})/\hbar$ requires an additional energy of $U>0$, leading to an effective repulsion of the second photon. This effect is similar to a Kerr-type nonlinearity of the form 
\begin{equation}
    H_{\textrm{Kerr}} = U_{\textrm{eff}}\,N(N-1)
    \label{eq:JC_kerr}
\end{equation} and therefore parallels the on-site interactions of the Bose-Hubbard model \cite{fisher1989boson, greentree2006quantum,schmidt2009strong}. However, this comparison must be approached with caution due to two key subtleties. First, applicability for an arbitrary number of excitations requires that $U_{\textrm{eff}}$ itself depends on the number operator $N$, as in the large excitation limit the Jaynes-Cummings ladder approaches a linear spectrum \cite{Raftery2014,Carmichael2015}. This idea -- defining an explicit, excitation number dependent $U_{\textrm{eff}}$ -- has been explored in several publications to date \cite{greentree2006quantum, koch2009superfluid, hartmann2016quantum} but, as noted in Ref. ~\onlinecite{koch2009superfluid}, leads to inaccuracies in the dispersive coupling regime. Second, the operators appearing in Eq. (\ref{eq:JC_kerr}) correspond not to the number of photons in the cavity, but the number of \emph{dressed} photons. As a result, the very nature of the underlying excitations themselves depend upon the parameter regime, changing from photonic in the dispersive regime to polaritonic for resonant coupling, behavior which is not apparent from Eq. (\ref{eq:JC_kerr}). A useful effective bosonic many-body representation of Eq. (\ref{eq:JC}) therefore requires a more careful consideration of these subtleties.

In the following subsections, we present a transformed representation of Eq. (\ref{eq:JC}) which makes explicit the bosonic many-body interactions generated through photon-blockade for general system parameters. In contrast with similar methods in the literature relying on Schrieffer-Wolff perturbation theory \cite{boissonneault2009dispersive, blais2004cavity, blais2020circuit}, our approach is applicable for both resonant ($\Delta\ll g$) and dispersive ($\Delta\gg g$) light-matter coupling. Through techniques of unitary transformation, we systematically develop an exact many-body description of the Jaynes-Cummings Hamiltonian and expose a hierarchy of normally ordered, effective $k$-body interactions and their parameter-dependent scaling. The end result is an exact generalization of Eq. (\ref{eq:JC_kerr}) which is absent of excitation number dependent coefficients. Particular attention is given in identifying the physically appropriate basis for the many-body interactions as it has been shown that insulator-to-superfluid quantum phase transitions of the JCH model are accompanied by a polaritonic-to-photonic transition in the nature of the excitations \cite{greentree2006quantum,hartmann2006strongly,Noh2016}.

\subsection{Unitary diagonalization of the Jaynes-Cummings Hamiltonian}\label{subsec:JC_diag}

While it is straightforward to find the eigenvalues and eigenvectors of the Jaynes-Cummings Hamiltonian by considering each excitation number manifold individually, an alternate route toward diagonalizing Eq. (\ref{eq:JC}) involves unitary transformation of the canonical operators. This approach was first reported in Ref.~\onlinecite{carbonaro1979canonical} and has since been adopted in a number of more recent works \cite{blais2004cavity, boissonneault2009dispersive, blais2020circuit}. At first glance, this strategy appears to be a more complicated pathway toward computing the well-known eigenvalues and eigenvectors of Eqs. (\ref{eq:JCeigenval}$-$\ref{eq:JCeigenvec}). However, it provides additional physical insight into the diagonal form of the Hamiltonian through an analytic understanding of the dressed canonical operators and will allow us to more clearly compare between Hamiltonians endowed with Jaynes-Cummings interactions and those having two-body bosonic interactions of the form of Eq. ($\ref{eq:JC_kerr}$).

We begin by writing the Jaynes-Cummings Hamiltonian as
\begin{equation}
H=H_0 + \hbar g I_+,
\label{eq:JC_short}
\end{equation} where we have adopted the shorthand notation \cite{boissonneault2009dispersive}
\begin{equation}
    \begin{split}
    H_0 &= \hbar\omega_c a^\dagger a+\frac{1}{2}\hbar\omega_a\sigma^z \\
        I_{\pm}&=a^\dagger \sigma^- \pm a \sigma^+.
    \end{split}
    \label{eq:Hamdefs}
\end{equation}
Defining the unitary transformation operator
\begin{equation}
    \mathcal{U} = e^{-\Lambda I_-},
    \label{eq:U}
\end{equation}
we aim to find the appropriate choice of $\Lambda$ for which the Hamiltonian is diagonal once cast in terms of the transformed operators $\widetilde{a}=\mathcal{U}^\dagger a\, \mathcal{U}$ and $\widetilde{\sigma}^-=\mathcal{U}^\dagger \sigma^- \mathcal{U}$. Due to the unitarity of $\mathcal{U}$, all commutation relations are invariant under transformation. 

Here we employ the method of \emph{explicit} transformation, whereby the Hamiltonian $H$ is rewritten in terms of transformed operators. This strategy typically entails finding closed analytic relationships between each canonical operator $\mathcal{O}$ and its transformed pair $\widetilde{\mathcal{O}}$ using the Baker-Campbell-Hausdorff formula \cite{wagner1986unitary},
\begin{equation}
    \begin{split}
     \widetilde{\mathcal{O}} &= \mathcal{U}^\dagger \mathcal{O}\mathcal{U}\\
     &=\mathcal{O} + [\mathcal{O},S] + \frac{1}{2!}[[\mathcal{O},S],S] + \frac{1}{3!}[[[\mathcal{O},S],S],S] +\ldots,
     \end{split}
     \label{eq:BakerCambellHaus}
\end{equation}
where $\mathcal{U}=e^S$. In the case of the Jaynes-Cummings Hamiltonian, however, direct application of Eq. (\ref{eq:BakerCambellHaus}) to the canonical operators $a$ and $\sigma^-$ leads to an infinite series of commutation relations which do not close, and a ``nonunitarian short circuit'' must be employed to obtain closed form expressions through this approach \cite{wagner1986unitary}. Instead, it is advantageous to transform $H_0$ and $I_+$ in their entirety. Using the commutation relations
\begin{equation}
    \begin{split}
        [H_0,I_-]&=-\hbar\Delta I_+\\
        [I_+,I_-]&=2 N \sigma^z \\
    \end{split}
    \label{eq:helpfulcomms}
\end{equation}
along with the inverted form of Eq. (\ref{eq:BakerCambellHaus}), it can be shown that 
\begin{equation}
    \begin{split}
        H_0 &= \widetilde{H}_0 + \Lambda[\widetilde{H}_0,I_-]+\frac{\Lambda^2}{2!}[[\widetilde{H}_0,I_-],I_-] + \ldots\\
        &=\widetilde{H}_0 - \hbar\Delta \sum_{n=1}\frac{\Lambda^n}{n!}\widetilde{F}_{n-1}  \\
        I_+ &= \widetilde{I}_+ + \Lambda[\widetilde{I}_+,I_-]+\frac{\Lambda^2}{2!}[[\widetilde{I}_+,I_-],I_-] + \ldots\\
        &=\sum_{n=0}\frac{\Lambda^n}{n!}\widetilde{F}_{n}
    \end{split}
    \label{eq:commexp}
\end{equation}
where $\widetilde{F}_n$ is the $n$th order commutator of $\widetilde{I}_+$ and $I_-$ given by
\begin{equation}
    \widetilde{F}_n=\begin{cases} 
      (-1)^{\frac{n-1}{2}} (2 \sqrt{N})^{n+1}\,\widetilde{\sigma}^z/2 & n \textrm{ odd} \\
      (-1)^{\frac{n}{2}} (2 \sqrt{N})^n\,\widetilde{I}_+ & n \textrm{ even} \\
   \end{cases}
   \label{eq:Fn}
\end{equation}
and transformed operators are indicated by tildes. Note that both $N$ and $I_-$ commute with $\mathcal{U}$ and, consequently, tildes on these operators are neglected for simplicity.

Using the relations in Eq. (\ref{eq:Fn}), the commutator expansions of $H_0$ and $I_+$ may be formally summed and substituted into Eq. (\ref{eq:JC_short}), yielding
\begin{equation}
    \begin{split}
        H &= \hbar\omega_c\left(N-\frac{1}{2}\right) \\&- \frac{\hbar}{2\sqrt{N}}\left[\Delta\sin(2\Lambda\sqrt{N})-2g\sqrt{N} \cos(2\Lambda\sqrt{N})\right]\widetilde{I}_+ \\
        &+\frac{\hbar}{2}\left[\Delta\cos(2\Lambda\sqrt{N})+2g\sqrt{N} \sin(2\Lambda\sqrt{N})\right]\widetilde{\sigma}^z.
    \end{split}
    \label{eq:JC_transf_before}
\end{equation}
Diagonalization is achieved through elimination of the second term proportional $\widetilde{I}_+$, leading to the constraint
\begin{equation}
    \Lambda(N) \equiv \frac{\theta(N)}{\sqrt{N}} = \frac{1}{2\sqrt{N}}\tan^{-1}\left(2\lambda\sqrt{N}\right),
    \label{eq:mixingangle}
\end{equation}
defined here in terms of the mixing angle $\theta$, previously introduced in Eq. (\ref{eq:JCeigenvec}) but now appearing as a function of the number operator $N$ rather than its eigenvalue $n$. Critically, $\Lambda$ is also a function of the operator $N$. This is allowed only because $N$ commutes with $H_0$ and $I_\pm$ and therefore may be effectively treated as a scalar in writing the commutation series of Eq. (\ref{eq:commexp}). We emphasize, however, that caution must be exercised in endowing $\Lambda$ with arbitrary operator dependence.

With the above choice of $\Lambda$, simplification of Eq. (\ref{eq:JC_transf_before}) yields
\begin{equation}
    H = \hbar\omega_c\left(N-\frac{1}{2}\right) + \frac{\hbar}{2}\Delta\sqrt{1+4\lambda^2N}\,\widetilde{\sigma}^z.
\label{eq:JCsqrt}
\end{equation}
The above Hamiltonian is now entirely diagonal written in terms of the dressed bosonic and TLS operators, the former appearing via the total number operator $N=\widetilde{a}^\dagger \widetilde{a} + \widetilde{\sigma}^+\widetilde{\sigma}^-$. While it is evidently clear that this Hamiltonian returns the same eigenvalues previously reported in Eq. (\ref{eq:JCeigenval}), this procedure allows for an exact operator representation of Jaynes-Cummings Hamiltonian in terms of the dressed \emph{operators} rather than a description of the dressed \emph{states} provided by the manifold-by-manifold approach. As will be shown in later sections, the dressed operator form of the Jaynes-Cummings Hamiltonian provides a deeper understanding of the underlying bosonic many-body interactions mediated by the TLS. More immediately, it is imperative to first understand how the dressed operators act on the composite Hilbert space of the dressed states of the Jaynes-Cummings Hamiltonian.

\subsection{Behavior of the dressed operators}\label{subsec:dressedops}
As previously discussed, direct transformation of the bosonic and TLS operators $a$ and $\sigma^-$ does not yield easily interpretable closed-form expressions for the dressed operators $\widetilde{a}$ and $\widetilde{\sigma}^-$. Despite this, one may still determine the action of the dressed operators on the eigenstates of Eq. (\ref{eq:JCeigenvec}) by transforming both the states and operators to the original basis where the action of the bare operators is known. Given the unitary transformation Eq. (\ref{eq:U}), the state $\ket{\Psi}$ transforms according to
\begin{equation}
    \ket{\Psi}_S = e^{-S}\ket{\Psi}
\end{equation}
where the subscript identifies a state transformed with respect to the generating function $S=-\Lambda I_-$. In order to work out transformations of the states explicitly, it is helpful to first cast the unitary operator $\mathcal{U}=e^{S}$ in an alternate form via Taylor expansion and subsequent resummation. In particular, it may be shown that
\begin{equation}
    e^{\pm S} = \cos(\theta) \mp \frac{1}{\sqrt{N}}\sin(\theta)I_-,
\end{equation}
where $\theta$ is the mixing angle defined in Eq. (\ref{eq:mixingangle}). Then the basis states $\left\{\ket{n,g}, \ket{n,e}\right\}$ transform as
\begin{equation}
    \begin{split}
        \ket{n,g}_S &= \ket{n,-} = \cos(\theta)\ket{n,g} - \sin(\theta) \ket{n-1,e} \\
        \ket{n-1,e}_S &= \ket{n,+}  =\sin(\theta)\ket{n,g} +\cos(\theta)\ket{n-1,e},
    \end{split}
    \label{eq:eigenvecs}
\end{equation}
where we have made explicit the equivalence between the transformed states and the well-known eigenstates of the Jaynes-Cummings Hamiltonian introduced in Eq. (\ref{eq:JCeigenvec}). As expected, then, the unitary operator $\mathcal{U}^\dagger$ maps the bare basis states onto the set  of eigenstates $\left\{\ket{n,g}_S, \ket{n,e}_S\right\}$. It is important to note that the ground state is included within the set $\left\{\ket{n,-}\right\}$ which corresponds to the ``lower branch'' of the Jaynes-Cummings ladder for $\Delta\geq 0^+$ ($\theta>0$) and to the ``upper branch'' for $\Delta\leq 0^-$ ($\theta<0$), where superscripts indicate the direction of approach for the case $\Delta=0$. We note, however, that the choice of which branch includes the ground state is arbitrary, and the roles of $\ket{n,-}$ and $\ket{n,+}$ may be reversed by adding an overall minus sign to $S$ or, equivalently, swapping $\mathcal{U}$ and $\mathcal{U}^\dagger$ in the convention adopted for the similarity transform Eq. (\ref{eq:BakerCambellHaus}). Nonetheless, a choice has been made in identifying $\ket{0,-}$ with the ground state and, because the $n=0$ manifold consists of only one state, $\ket{0,+}$ does not represent a physical state of the system.

\begin{figure}
\includegraphics[width=0.35\textwidth]{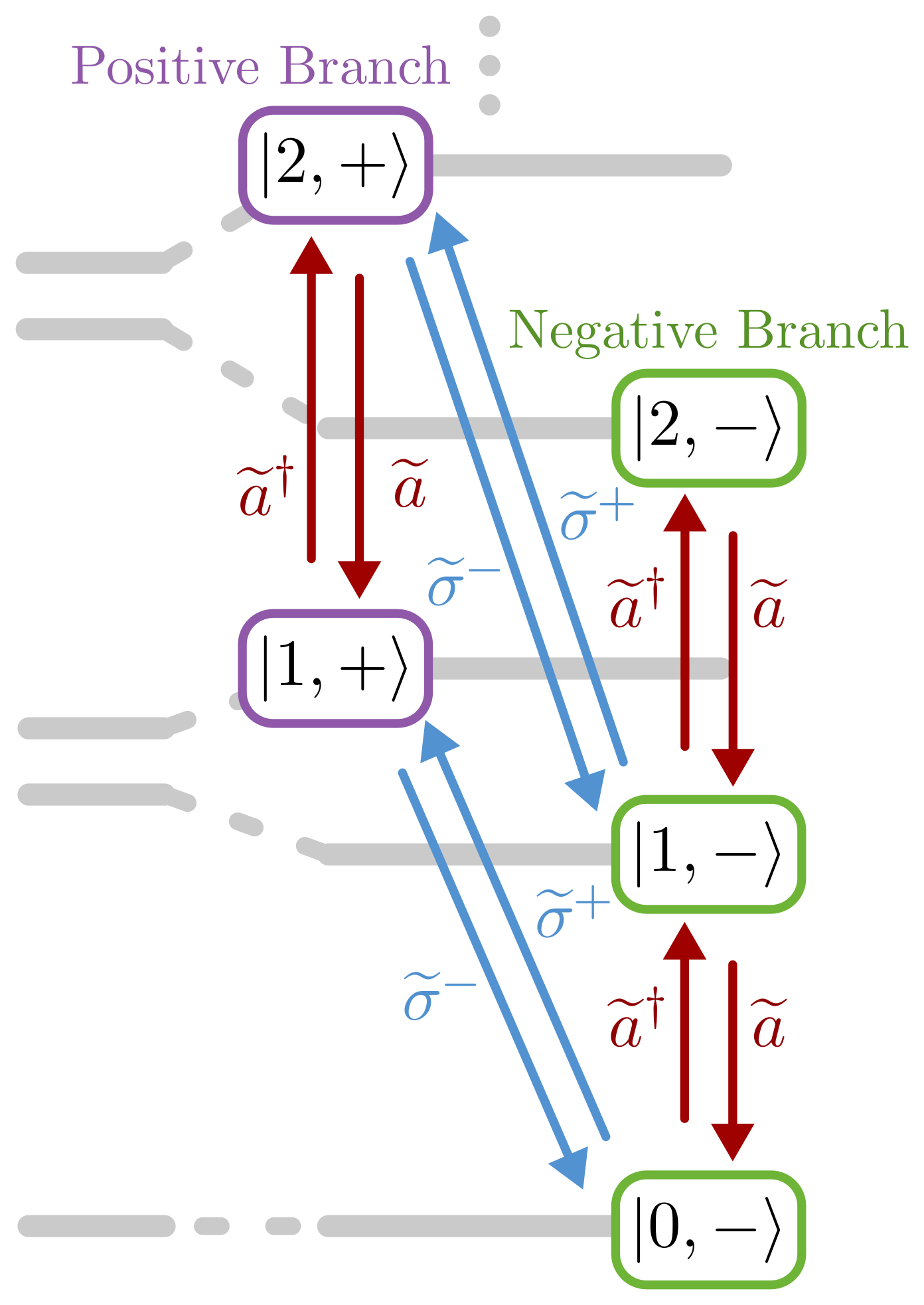}
\caption{\label{fig:f2} Action of the dressed operators on the Jaynes-Cummings ladder. Upon diagonalization, the Jaynes-Cummings Hamiltonian may be repackaged into a positive and negative branch, the former comprising the states $\ket{n,-}$ and the latter $\ket{n,+}$. The dressed bosonic operators $\widetilde{a}^\dagger$ and $\widetilde{a}$ induce transitions between states of the same branch, raising and lowering the total number of excitations by one, respectively. In contrast, the dressed psuedospin operators $\widetilde{\sigma}^+$ and $\widetilde{\sigma}^-$ facilitate transitions between the two branches. Similar to the bare pseudospin operators $\sigma^+$ and $\sigma^-$ acting on the states $\ket{e}$ and $\ket{g}$, respectively, applying $\widetilde{\sigma}^{\pm}$ to a state denoted by the same sign returns zero.}
\end{figure}

Turning now to the action of the dressed operators, one may show that for a general operator $\mathcal{O}$,
\begin{equation}
        \widetilde{\mathcal{O}}\ket{n,m}_S =  e^{-S}\mathcal{O}\ket{n,m}=  (\mathcal{O}\ket{n,m})_S \\
\label{eq:opaction}
\end{equation}
where $m=\{g,e\}$. Accordingly, the action of the operator $\widetilde{\mathcal{O}}$ in the basis of transformed states $\ket{n,m}_S$ is exactly analogous the action of $\mathcal{O}$ in the original basis spanned by the Fock states $\ket{n,m}$. The action of the dressed operators on the conventionally labeled states $\ket{n,\pm}$, however, is more subtle as here $n$ indicates the excitation manifold or, equivalently, the total number of combined bosonic and TLS excitations rather than the number of dressed bosonic excitations alone as in the labeling $\ket{n,m}_S$. We emphasize that these subtleties are solely a consequence of notation and are of little physical importance, and as a result it is often simpler to work with the more physically apparent notation $\ket{n,m}_S$ labeling Fock states in the dressed boson/TLS basis. Still the action of the dressed operators on the states $\ket{n,\pm}$ may be easily worked out through combination of Eqs. (\ref{eq:eigenvecs}$-$\ref{eq:opaction}), with results summarized for reference in Table \ref{tab:ops} and Fig. \ref{fig:f2}.

Although the description of the dressed operators thus far has been exact for general system parameters, it is instructive to contrast two important parameter regimes of the Jaynes-Cummings model: resonant coupling ($\lambda\gg 1$) and dispersive coupling ($\lambda\ll 1$). In the former case, the mixing angle $\theta$ approaches $\pm\pi/4$ and the eigenvectors of Eq. (\ref{eq:eigenvecs}) are maximally mixed superpositions of bosonic cavity and atomic excitations. Consequently, the dressed bosonic and TLS operators induce transitions between the hybridized light-matter eigenstates of the system and the fundamental excitations of the system are polaritonic. In contrast, the dispersive regime is most easily analyzed by first recognizing that Taylor expansion of the rightmost side of Eq. (\ref{eq:mixingangle}) yields $\Lambda\approx \lambda$ and therefore the unitary transformation operator may be approximated as $\mathcal{U}=e^{-\Lambda I_-}\approx e^{-\lambda I_-}$. Approximate forms of the transformed operators are then obtained through Schrieffer-Wolff perturbation theory for $\lambda\ll 1$ \cite{boissonneault2009dispersive, blais2020circuit}, leading to
\begin{equation}
    \begin{split}
        \widetilde{a} &\approx a - \lambda \sigma^- \\
        \widetilde{\sigma}^- &\approx \sigma^- - \lambda a\sigma^z,
    \end{split}
    \label{eq:ops_approx_disp}
\end{equation}
where only first order corrections in $\lambda$ have been kept. Here, the bosonic operators $\widetilde{a}^\dagger$ and $\widetilde{a}$ create and destroy photons weakly perturbed by the presence of the TLS. Likewise, the perturbed operators $\widetilde{\sigma}^+$ and $\widetilde{\sigma}^-$ include the expected action of raising or lowering the bare TLS and additionally inherit a small photonic contribution conditioned on the state of the bare TLS via $\sigma^z$.

It is important to note that the transformed operators $\widetilde{a}$ and $\widetilde{\sigma}^-$ form an appropriate operator basis regardless of the parameter regime, and their action on the transformed states is independent of whether the system is resonantly or dispersively coupled. However, the underlying character of the transformed operators and states changes as a function of system parameters, most easily seen by relating the transformed operators and states back to those describing the uncoupled system as shown above. For example, it is clear that the bosonic operators $\widetilde{a}$ and $\widetilde{a}^\dagger$ describe either creation and annihilation of polaritons or photons depending on the value of the mixing angle $\theta$ (or equivalently, $\lambda$). As a result, the dressed operator description of the Jaynes-Cummings Hamiltonian is appropriate independent of the parameter regime under consideration. Still, it is crucially important to maintain an understanding of the parameter-dependent underlying physical character of the excitations described by the transformed operators and states. This will hold especially true in Section \ref{sec:2JCH} where it will be shown that the physical interpretation of the distinct quantum phases of a two-site JCH model requires knowledge of the underlying nature of the transformed states across parameter space.

\setlength\tabcolsep{5pt}
\begin{table}
  \centering
    \begin{tabular}{ |c||r|r|  }
     \hline
      & \multicolumn{1}{|c|}{$\ket{n,-}=\ket{n,g}_S$} & \multicolumn{1}{|c|}{$\ket{n,+}=\ket{n-1,e}_S$}\\
      \hline
     \,$\widetilde{a}$\,           &$\sqrt{n}\ket{n-1,-}$     & $\sqrt{n-1}\ket{n-1,+}$\\
     \,$\widetilde{a}^\dagger$\,    &$\sqrt{n+1}\ket{n+1,-}$   & $\sqrt{n}\ket{n+1,+}$\\
     \,$\widetilde{a}^\dagger\widetilde{a}$\,          &$n\ket{n,-}$              & $(n-1)\ket{n,+}$\\
     \,$\widetilde{\sigma}^-$\,    &0                           & $\ket{n-1,-}$\\
     \,$\widetilde{\sigma}^+$\,    &$\ket{n+1,+}$             & 0\\
     \,$\widetilde{\sigma}^z$\,    &$-\ket{n,-}$              & $\ket{n,+}$\\
     \,$\widetilde{\sigma}^+\widetilde{\sigma}^-$\,    &0            & $\ket{n,+}$\\
     \,$\widetilde{\sigma}^-\widetilde{\sigma}^+$\,    &$\ket{n,-}$              &0\\
     \,$N$\,                                           &$n\ket{n,-}$              & $n\ket{n,+}$\\

     \hline
    \end{tabular}
    \caption{\label{tab:ops} Behavior of the dressed operators acting on the Jaynes-Cummings ladder states $\ket{n,\pm}$.}
\end{table}

\subsection{Revealing the hidden many-body nature of the Jaynes-Cummings Hamiltonian}\label{subsec:manybodyrep}

Paired with the results of the previous section, the Hamiltonian of Eq. (\ref{eq:JCsqrt}) provides a complete description of the Jaynes-Cummings Hamiltonian in the dressed operator basis. In its present form, the second term clearly endows the system with a nonlinear dependence on the total number of excitations, reminiscent of the Kerr-like, two-body bosonic interactions of the Bose-Hubbard model in Eq. (\ref{eq:JC_kerr}). The goal of this section is to make this analogy more apparent by casting Eq. (\ref{eq:JCsqrt}) in a form which accentuates the underlying many-body bosonic interactions. One route for achieving this involves Taylor expansion of Eq. (\ref{eq:JCsqrt}) about small values of
$\lambda$ and truncating at finite order \cite{boissonneault2009dispersive}. Alternatively, identical results are attained by direct Schrieffer-Wolff transformation of the Jaynes-Cummings Hamiltonian in its original representation, whereby the unitary operator $\mathcal{U}=e^{-\Lambda I_-}$ is replaced by its approximate form $\mathcal{U}=e^{-\lambda I_-}$ \cite{boissonneault2009dispersive, blais2020circuit} and all transformations are carried out to finite order. However, the two described strategies are only applicable in the dispersive limit where $\lambda\ll 1$, and it is therefore the purpose of this section to leverage the exact solution of Eq. (\ref{eq:JCsqrt}) toward a non-perturbative method equally applicable in both the dispersive and resonant coupling regimes.

Focusing on the nonlinear portion of Eq. (\ref{eq:JCsqrt}) alone, it is useful to define the function
\begin{equation}
    f(x) = \sqrt{1+4\lambda^2x}
\end{equation}
such that the dressed operator representation of the Jaynes-Cummings may be written as
\begin{equation}
    H = \hbar\omega_c\left(N-\frac{1}{2}\right)+\frac{\hbar}{2}\Delta f(N)\widetilde{\sigma}^z.
\end{equation} 
Using the identity $\widetilde{\sigma}^z = \widetilde{\sigma}^+\widetilde{\sigma}^- - \widetilde{\sigma}^-\widetilde{\sigma}^+$ and defining the projection operator $\mathcal{P}_{n}^\pm = \ket{n,\pm} \bra{n,\pm}$, the product $f(N)\widetilde{\sigma}^z$ may be further reexpressed as
\begin{equation}
    f(N)\widetilde{\sigma}^z = - \mathcal{P}_0 + \sum_{n=1}f(n)(\mathcal{P}_n^+-\mathcal{P}_n^-).
    \label{eq:fNsig_0}
\end{equation}
As shown in Appendix \ref{sec:App1}, one may Taylor expand $f(n)$ about $n=n_0$ and recast in terms of dressed operators to find
\begin{equation}
     \begin{split}
        f(N)\widetilde{\sigma}^z = &\sum_{r=0}^\infty\sum_{m=0}^r \binom{1/2}{r}\binom{r}{m}(2\lambda)^{2r} f(n_0)^{1-2r}(-n_0)^{r-m} \\
        \times&\sum_{k=0}^m(\widetilde{a}^\dagger)^k(\widetilde{a})^k \left[\bracenom{m+1}{k+1}\widetilde{\sigma}^+\widetilde{\sigma}^- - \bracenom{m}{k}\widetilde{\sigma}^-\widetilde{\sigma}^+\right],
     \end{split}
    \label{eq:fNsig_1}
\end{equation}
where $\bracenom{n}{k}$ are Stirling numbers of the second kind. Taking care to adjust upper and lower bounds as needed, the three sums appearing in Eq. (\ref{eq:fNsig_1}) may be reordered such that the total Hamiltonian becomes
\begin{equation}
    H = \hbar\omega_c\left(N-\frac{1}{2}\right) + \sum_{k=0}^{\infty}\frac{1}{k!}\left[ C^+_k\widetilde{\sigma}^+\widetilde{\sigma}^- + C^-_k\widetilde{\sigma}^-\widetilde{\sigma}^+\right](\widetilde{a}^\dagger)^k(\widetilde{a})^k,
\label{eq:Hfinal}
\end{equation}
where the coefficients of the $k$-body terms include the remaining sums over $m$ and $r$ in Eq. (\ref{eq:fNsig_1}). After partial resummation and further manipulation (see Appendix \ref{sec:App1}), it may be shown that these $k$-body interaction coefficients are given by
\begin{equation}
    \begin{split}
        C_k^-/\hbar &= -\frac{\Delta}{2}\sum_{p=0}^k\binom{k}{p}(-1)^{k+p}\sqrt{1+4\lambda^2p} \\
        C_k^+/\hbar &= \frac{\Delta}{2}\sum_{p=0}^k\binom{k}{p}(-1)^{k+p}\sqrt{1+4\lambda^2(p+1)}.
    \end{split}
    \label{eq:mbcoefficients}
\end{equation}
Together, Eqs. (\ref{eq:Hfinal} -- \ref{eq:mbcoefficients}) form an exact bosonic many-body representation of the Jaynes-Cummings Hamiltonian and constitute one of the primary results of this manuscript. Critically, this final form of the Hamiltonian is independent of the expansion point $n_0$. We note that
\begin{equation}
    \begin{split}
        \frac{1}{k!}(\widetilde{a}^\dagger)^k(\widetilde{a})^k\ket{n,-} &= \binom{n}{k}\ket{n,-} \\
        \frac{1}{k!}(\widetilde{a}^\dagger)^k(\widetilde{a})^k\ket{n,+} &= \binom{n-1}{k}\ket{n,+}
    \end{split}
\end{equation}
and thus each $k$-body term scales as $C_k^{\pm}$ multiplied by a combinatorial factor. When applied to the eigenstates $\ket{n,\pm}$, the infinite sum of $k$-body interactions may be evaluated, resulting in the closed form
\begin{equation}
    \sum_{k=0}^{\infty}\frac{1}{k!}C_k^\pm(\widetilde{a}^\dagger)^k(\widetilde{a})^k\ket{n,\pm} = \pm\frac{\hbar}{2}\Delta\sqrt{1+4\lambda^2n}\ket{n,\pm},
    \label{eq:sumkcheck}
\end{equation}
thus verifying that the dressed operator many-body form of the Jaynes-Cummings Hamiltonian in Eq. (\ref{eq:Hfinal}) returns the well known eigenvalues in Eq. (\ref{eq:JCeigenval}).

Critical to the usefulness of Eq. (\ref{eq:Hfinal}) is a clear partitioning of the Hilbert space into two branches, each spanned by either set of states $\ket{n-1,e}_S =\ket{n,+}$ or $\ket{n,g}_S =\ket{n,-}$. Because the two branches are uncoupled, one may consider each subspace independently. As previously discussed, it is the latter set which includes the global ground state $\ket{0,g}_S=\ket{0,-}$, and we thus focus our analysis on the ``negative'' branch, noting that much of the discussion follows similarly for the ``positive'' branch with the caveat that, there, the state $\ket{0,e}_S = \ket{1,+}$ effectively serves as the ground state within the subspace spanned by $\ket{n,+}$. We reemphasize, however, that the states $\ket{n,-}$ are the lower energy eigenstates of each excitation number manifold for $\Delta\geq0^+$ ($\theta>0$) only, and the eigenstates $\ket{n,-}$ exceed $\ket{n,+}$ in energy for $\Delta\leq0^-$ ($\theta<0$). As a result, one may access the entirety of the Jaynes-Cummings ladder for the resonant coupling case simply by choosing to approach $\Delta=0$ either from the positive or negative direction, yielding $\theta=\pi/4$ or $\theta=-\pi/4$, respectfully. We will find that this freedom allows for a mathematical description of either repulsive or attractive many-body interactions within the subspace of states $\ket{n,-}$ depending on the sign of $\theta$. Separately, in the dispersive regime, the negative branch comprises perturbed photonic excitations with the weakly dressed TLS in its unexcited state.

Before proceeding with a closer analysis of the coefficients $C_k^\pm$, it is important to note that the effects of environmental coupling have, up to this point, not been considered. As a result, the many-body terms of Eq. (\ref{eq:Hfinal}) seemingly play an important role for all $C_k^\pm\neq0$ and, as illustrated in Fig. (\ref{fig:f1}b), perfect photon blockade is achieved as long as $g\neq0$. In an experimental setting, however, coupling to the environment broadens the levels of the Jaynes-Cummings ladder such that photon blockade is impaired when the dominant rate of dissipation $\Gamma = \textrm{max}\{\kappa,\gamma\}$ exceeds the light-matter coupling strength $g$, where $\kappa$ and $\gamma$ denote the cavity and atomic linewidth, respectively. As a consequence, strong effective many-body interactions are realizable only in the strong coupling regime (i.e., $g>\Gamma$), as the impact of each $k$-body term depends not on $C_k^\pm$ alone, but rather on the ratio $C_k^\pm/\hbar\Gamma$. Although the effects of environmental coupling will not be explicitly considered in the present work, given the discussion above it is convenient to consider all parameters in units of $\Gamma$ as it determines the appropriate time scale for a specific realization of the Jaynes-Cummings Hamiltonian, allowing for a general discussion agnostic of the particulars of each experimental platform.

While Eqs. (\ref{eq:Hfinal}--\ref{eq:mbcoefficients}) provide an exact bosonic many-body representation of the Jaynes-Cummings Hamiltonian for general system parameters, the infinite sum over competing $k$-body terms obscures simple interpretation. It is therefore advantageous to closely analyze several limiting cases to gain insight into the contributions of the hierarchy of many-body terms appearing in Eq. (\ref{eq:Hfinal}). In the following, we restrict analysis to the few excitation limit and investigate both the dispersive and resonant coupling regimes independently. We then conclude the current section with a brief discussion of the more general $n$ excitation case.

\subsubsection{The few excitation limit: $n\leq 2$}
We begin by examining the Hamiltonian in Eq. (\ref{eq:Hfinal}) in the limit where the total number of excitations is fixed to two or fewer. In this scenario, the normally ordered terms $(\widetilde{a}^\dagger)^k(\widetilde{a}^\dagger)^k$ do not contribute for $k>3$ for the negative branch and $k>2$ for the positive branch. Consequently, the $n\leq 2$ limit allows for analysis of the Hamiltonian in the scenario where the highest order contributing many-body interactions correspond to two-body terms, leading to the effective Hamiltonian
\begin{equation}
    \begin{split}
        H_{n\leq 2}^{\textrm{eff}} =& \widetilde{\sigma}^+\widetilde{\sigma}^-\left[(\hbar\omega_c+C_1^+)\widetilde{a}^\dagger\widetilde{a} + \frac{1}{2}\hbar\omega_c + C_0^+\right] \\ 
        +& \widetilde{\sigma}^-\widetilde{\sigma}^+\left[(\hbar\omega_c+C_1^-)\widetilde{a}^\dagger\widetilde{a} + \frac{C_2^-}{2}\widetilde{a}^\dagger\widetilde{a}^\dagger\widetilde{a}\widetilde{a} - \frac{1}{2}\hbar\omega_c + C_0^-\right],
    \end{split}
    \label{eq:Heff_nleq2}
\end{equation}
where the first and second lines correspond to the effective Hamiltonian projected onto the positive and negative branches, respectively, and
\begin{equation}
    \begin{split}
        C_0^-/\hbar &= -\frac{\Delta}{2} \\
        C_1^-/\hbar &= -\frac{\Delta}{2}(-1+\sqrt{1+4\lambda^2})\\
        C_2^-/\hbar &= +\frac{\Delta}{2}(-1+2\sqrt{1+4\lambda^2} - \sqrt{1+8\lambda^2})\\
        C_0^+/\hbar &= -\frac{\Delta}{2}(\sqrt{1+4\lambda^2})\\
        C_1^+/\hbar &= -\frac{\Delta}{2}(-\sqrt{1+4\lambda^2}+\sqrt{1+8\lambda^2})\\
    \end{split}
\end{equation}
are the explicit forms of the interaction coefficients. In all cases, the above coefficients are written in such a way that the factor in parenthesis is positive for all values of $\lambda$ and therefore the overall sign of the coefficient is indicated explicitly in the prefactor. Notably, the overall sign of the coefficients $C_i^\pm$ depends upon the sign of the detuning $\Delta$. Two-body bosonic interaction terms appear only for the negative branch as the positive branch consists of states with the dressed TLS in its excited state, and limiting the total number of excitations to two or fewer therefore ensures at most one dressed photonic excitation. 

Focusing only on the negative branch, the Hamiltonian may be written within this subspace as
\begin{equation}
    H_{n\leq2}^{\textrm{eff}(-)} = (\hbar\omega_c+C_1^-)N + \frac{C_2^-}{2}N(N-1) - \frac{1}{2}\hbar\omega_c + C_0^-.
    \label{eq:Heff_nleq2_negbranch}
\end{equation}
where we have used the fact that $N$ and $\widetilde{a}^\dagger\widetilde{a}$ are identical for the negative branch. This effective Hamiltonian is, up to an overall energy shift, identical in form to the on-site terms of the Bose-Hubbard model \cite{greentree2006quantum},
\begin{equation}
    H_{\textrm{BH,on-site}} = -\mu N + \frac{U}{2}N(N-1),
    \label{eq:bosehubbard}
\end{equation}
where the on-site interaction strength $U$ is determined by $C_2^-$ and $N$ describes the number of $\emph{dressed}$ bosonic excitations. Despite the fact that the linear energy $\hbar\omega_c + C_1^-$ is strictly positive for realistic parameters and thus naturally describes a system with $\mu<0$, we note that one may transform to a rotating frame via the unitary operator $e^{-i\omega_c \widetilde{a}^\dagger\widetilde{a}t}$ such that $C_1^-$, which is negative for $\Delta\geq 0^+$, becomes analogous to the chemical potential. As discussed in Section {\ref{subsec:dressedops}}, the dressed operators $\widetilde{a}^\dagger$ and $\widetilde{a}$ describe creation and annihilation of bosonic excitations whose character varies from polaritonic ($\lambda\gg1$) to photonic ($\lambda\ll1$) depending on the choice of $g$ and $\Delta$. Furthermore, as the overall sign of $C_2^-$ is determined by the sign of $\Delta$, the interaction energy $U$ can be either positive or negative. The former case results in an effective polariton-polariton (or photon-photon) repulsion, whereas the latter corresponds to polariton-polariton (or photon-photon) attraction.

\begin{figure}
\includegraphics[width=0.45\textwidth]{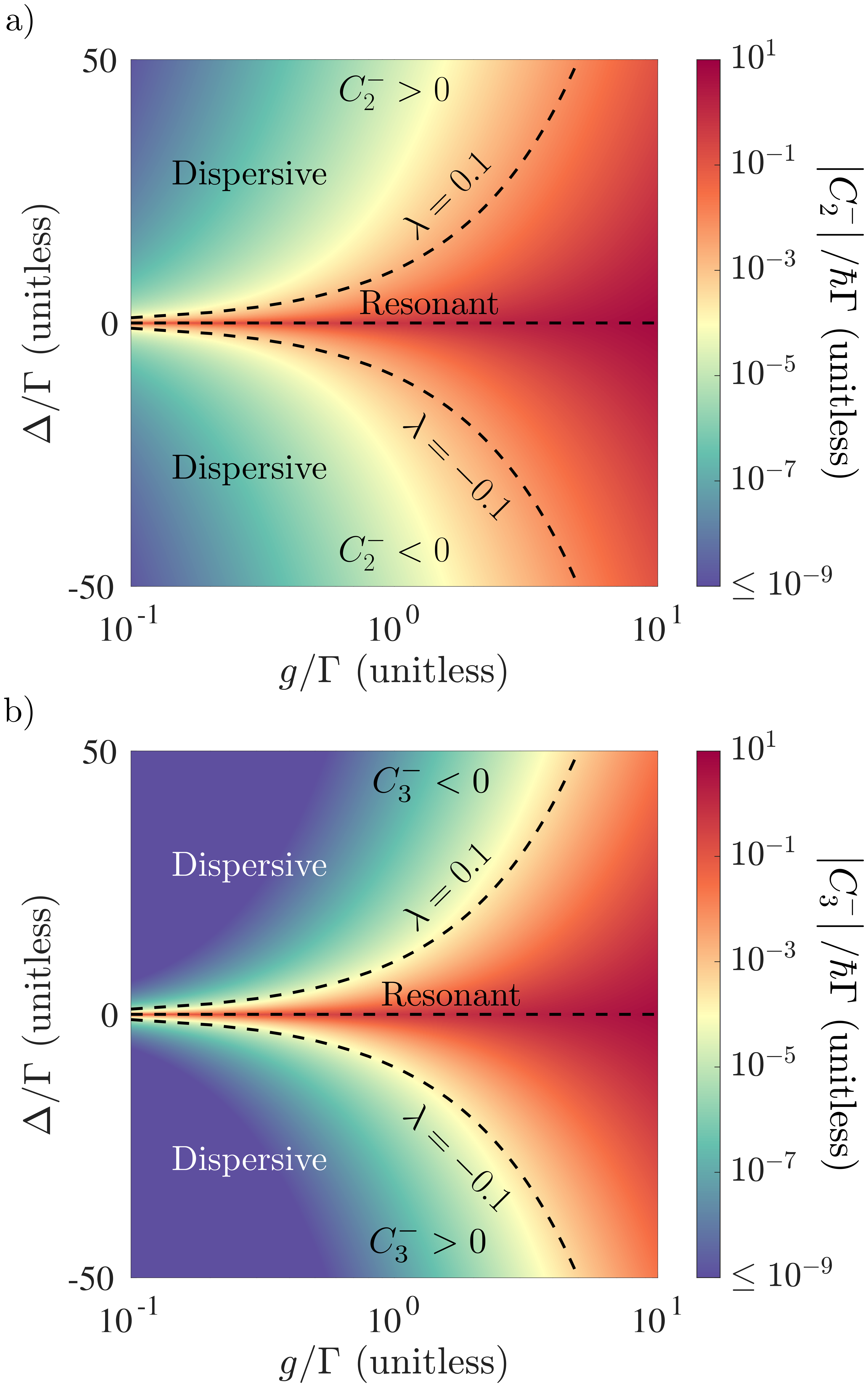}
\caption{\label{fig:f3} (a) Sign and magnitude of the two-body coefficient $C_2^-$ as a function of $\Delta$ and $g$. For practical purposes, all parameters are normalized to $\Gamma$ which sets the relevant frequency scale for the particular experimental platform in consideration. Regardless of the parameter regime, $C_2^-$ always takes the same sign as $\Delta$. As discussed in the main text, $C_2^-$ is discontinuous at $\Delta=0$ and may take on a positive or negative values depending on the direction of approach. For all $g$, $C_2^-$ is maximized for resonant coupling ($\lambda  \gg 1$), and is comparably smaller for dispersive coupling ($\lambda\ll 1$), requiring $g/\Gamma\gtrsim 1/2\lambda^3$ to achieve strong photon-photon interactions ($C_2^-/\Gamma\gtrsim 1$). The dashed line indicates $\lambda=0.1$, typically considered the onset of the dispersive regime. (b) Sign and magnitude of the three-body coefficient $C_3^-$ as a function of $\Delta$ and $g$, all normalized to $\Gamma$. $C_3^-$ displays a qualitatively similar trend to $C_2^-$, taking on a maximal magnitude at $\Delta=0$ and falling off rapidly for decreasing $\lambda$. For all parameters, $|C_3^-|<|C_2^-|$ and, in contrast to $C_2^-$, the sign of $C_3^-$ is opposite to that of $\Delta$.}
\end{figure}

Fig. \ref{fig:f3}a shows the absolute value of $C_2^-$ as a function of system parameters $g$ and $\Delta$, all relative to a fictitious dissipation rate $\Gamma$ which sets the relevant energy scale pertaining to a particular experimental platform, as discussed previously. As expected, the scaling of the two-body interaction is largest for resonant coupling where the bosonic modes and TLS maximally mix. Evaluating $C_2^-$ for the perfectly resonant case leads to
\begin{equation}
    C_2^-/\hbar=\pm(2-\sqrt{2})g, \qquad (\textrm{resonant})
    \label{eq:C2m_res}
\end{equation}
where the sign of $C_2^-$ is determined by the direction in which $\Delta=0$ is approached and it is assumed that $g\geq 0$. Then for resonant coupling, $C_2^-$ scales linearly with $g$ and strong two-body interactions ($C_2^-\gtrsim \hbar\Gamma$) are achieved for 
\begin{equation}
    g/\Gamma \gtrsim 1/(2-\sqrt{2}),
\end{equation}
a slightly higher threshold than strong coupling.

In the dispersive regime, $C_k^-$ depends nonlinearly on $\lambda$ and, as a result, spans many orders of magnitude for constant $\Delta$ depending upon the value of $g$. Expanding $C_2^-$ about small values of $\lambda$ leads to the result
\begin{equation}
    C_2^-/\hbar \approx 2\lambda^3g, \qquad (\textrm{dispersive})
    \label{eq:C2m_dis}
\end{equation}
indicating that the fall off of two-body interactions for decreasing $g$ is dependent on how far into the dispersive regime the system is tuned. To realize strong two-body interactions (i.e., $C_2^-\gtrsim\hbar\Gamma$) in the dispersive coupling regime, exceptionally large values of $g/\Gamma$ must be attained such that the condition
\begin{equation}
    g/\Gamma\gtrsim1/2\lambda^3
    \label{eq:stronginter_cond}
\end{equation}
is satisfied, a limit which has been approached in circuit QED platforms (for $\lambda\sim0.1$), reaching values of $g/\Gamma$ in the several hundreds \cite{Schoelkopf2008}. Eq. (\ref{eq:stronginter_cond}) may be thought of as a higher order generalization of the strong-dispersive regime \cite{Gambetta2006,Schuster2007}, defined by the condition $g/\Gamma\gtrsim1/\lambda$ (for $\lambda\lesssim 0.1$) which characterizes the portion of parameter space in which the first order frequency shift $C_1^{\pm}$ exceeds $\hbar\Gamma$.

While boson-boson interactions are most easily attained in the case of resonant coupling, it is in the dispersive parameter regime in which the bosonic many-body interactions take on a photonic nature. As we shall see in Sec. \ref{sec:2JCH} where the present analysis is extended to a two-site Jaynes-Cummings-Hubbard system, it is \emph{photonic} two-body interactions in the dispersive regime, rather than \emph{polaritonic} two-body interactions on resonance, which will mostly clearly provide a route for analog quantum simulation of Bose-Hubbard physics.


\subsubsection{The few excitation limit: $n\leq 3$}

The Hamiltonian in Eq. (\ref{eq:Heff_nleq2}) is exact for $n\leq2$. Consideration of states with $n=3$ requires inclusion of three-body terms, leading to the effective Hamiltonian
\begin{equation}
    \begin{split}
        H_{n\leq 3}^{\textrm{eff}} = H_{n\leq 2}^{\textrm{eff}} &+ \widetilde{\sigma}^+\widetilde{\sigma}^-\left[ \frac{C_2^+}{2}\widetilde{a}^\dagger\widetilde{a}^\dagger\widetilde{a}\widetilde{a}\right] \\
        &+\widetilde{\sigma}^-\widetilde{\sigma}^+\left[ \frac{C_3^-}{2}\widetilde{a}^\dagger\widetilde{a}^\dagger\widetilde{a}^\dagger\widetilde{a}\widetilde{a}\widetilde{a}\right],
    \end{split}
\end{equation}
where
\begin{equation}
    \begin{split}
        C_3^-/\hbar &= -\frac{\Delta}{2}(-1 + 3\sqrt{1+4\lambda^2} - 3\sqrt{1+8\lambda^2}+\sqrt{1+12\lambda^2}) \\ 
        C_2^+/\hbar &= +\frac{\Delta}{2}(-\sqrt{1+4\lambda^2} + 2\sqrt{1+8\lambda^2} - \sqrt{1+12\lambda^2}) \\ 
    \end{split}
\end{equation}
describe the strength of three-body (two-body) interactions within the negative (positive) branch of the Jaynes-Cummings ladder. Similar to the the $n\leq2$ case, three-body terms do not appear for the positive branch as the states considered allow for up to two bosonic excitations. We note that the trends followed by the positive branch for $n\leq3$ are similar to those of the negative branch for $n\leq2$ (with signs reversed), and therefore will not be explicitly discussed. 

Fig. \ref{fig:f3}b shows the magnitude of $C_3^-$ as a function of $\Delta$ and $g$, again relative to the  maximal dissipative rate $\Gamma = \textrm{max}\{\kappa,\gamma\}$. Notably, $C_2^-$ and $C_3^-$ differ by an overall sign with the latter of smaller magnitude for all parameters. Otherwise the two follow a similar trend, albeit with $C_3^-$ declining much more rapidly with decreasing $g$.

Following the analysis of the two-body interaction strength $C_2^-$, it is helpful to derive expressions for $C_3^-$ for the cases of resonant and dispersive coupling. For the former, evaluating $C_3^-$ for $\Delta=0$ leads to
\begin{equation}
    C_3^- = \mp(3-3\sqrt{2}+\sqrt{3})g \qquad (\textrm{resonant}),
\end{equation}
where, similar to Eq. (\ref{eq:C2m_res}), the sign of $C_3^-$ is dependent on the direction in which $\Delta=0$ is approached and the overall expression is proportional to $g$, here with a smaller prefactor such that $|C_2^-|>|C_3^-|$.

In contrast, evaluating $C_3^-$ for small values of $\lambda$ via Taylor expansion yields
\begin{equation}
    C_3^- \approx -12\lambda^5g \qquad (\textrm{dispersive}),
\end{equation}
similar to the result Eq. (\ref{eq:C2m_dis}) yet scaling at fifth order in $\lambda$ rather than third. Consequently, $C_3^-$ falls off much more rapidly than $C_2^-$ in the dispersive regime, indicating that the strength of three-body interactions are small relative to their two-body counterparts and may therefore be discarded for small enough $\lambda$. For all $g$ and $\Delta$, $C_2^-$ and $C_3^-$ are of opposite sign and therefore counteract one another in systems with at least three excitations, with positively and negative valued interactions describing repulsion and attraction, respectively.


\subsubsection{Nature of the many-body coefficients for arbitrary $n$}
Following the preceding analysis of the parameter dependent strength of two- and three-body interactions in the few excitation limit, extension to the general $n$ excitation limit is straightforward. Focusing again on the negative branch, it is convenient to independently analyze the form of the $k$-body coefficient $C_k^-$ for the cases of resonant and dispersive coupling. It is worth emphasizing again that for any finite $n$, each $k$-body term will only contribute if $k\leq n$, and the sum in Eq. (\ref{eq:Hfinal}) therefore always terminates. However, the $n\rightarrow \infty$ limit of the Jaynes-Cummings Hamiltonian is important to analyze as the eigenspectrum becomes approximately linear, inhibiting photon blockade for large values of $n$ \cite{Raftery2014, Carmichael2015}. Evaluating the general form of $C_k^-$ for $\Delta=0$ (see Eq. (\ref{eq:mbcoefficients})), we find
\begin{equation}
    C_k^- = \pm(-1)^k\left[\sum_{p=1}^k\binom{k}{p}(-1)^{p+1}\sqrt{p}\right]g \qquad (\textrm{resonant}),
    \label{eq:Ckm_res}
\end{equation}
where the upper and lower signs corresponds to the limit $\Delta\rightarrow 0^\pm$ and $\theta=\pm\pi/4$. Therefore the linear relationship with $g$ previously found for $C_2^-$ and $C_3^-$ is general for all $k$. Furthermore, the factor in parentheses is positive and convergent for all $k$. The overall sign of the coefficients $C_k^-$ therefore alternate in $k$, a trend which can be shown more generally from Eq. (\ref{eq:mbcoefficients}) without specializing to the case of resonant coupling. In the limit of very large $k$, the above sum asymptotically trends toward the closed expression \cite{harvardProf} $C_k^-=(-1)^k g/\sqrt{\pi\ln(k)}$ and thus vanishes in the limit $k\rightarrow\infty$.

For the case of dispersive coupling ($\lambda\ll1$), $C_k^-$ may be written as 
\begin{equation}
    C_k^-/\hbar \approx -k!\binom{1/2}{k}(2\lambda)^{2k-1}g \qquad (\textrm{dispersive}),
    \label{eq:Ckm_dis}
\end{equation}
where only the lowest order term in $\lambda$ has been retained.
In deriving this expression, $\sqrt{1+4\lambda^2p}$ was evaluated using a binomial expansion which, strictly speaking, is convergent only for $p\leq k<1/4\lambda^2$, setting an upper bound of $\lambda < \sqrt{1/4k}$ for which Eq. (\ref{eq:Ckm_dis}) is valid.

\begin{figure*}
\includegraphics[width=0.7\textwidth]{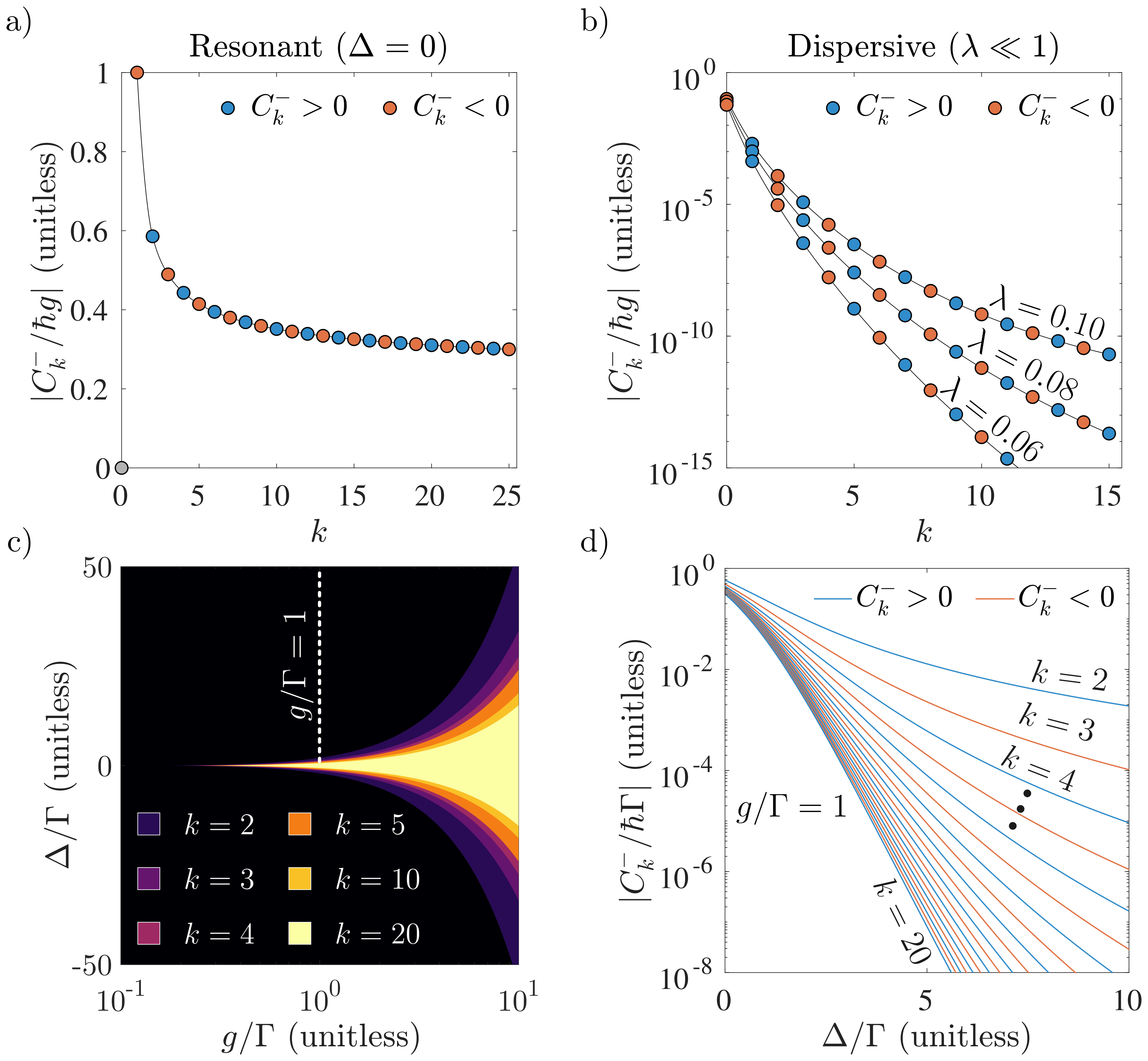}
\caption{\label{fig:f4} Top panels: sign and magnitude of the $k$-body coefficient $C_k^-$ normalized to $\hbar g$ for (a) resonant and (b) dispersive coupling, plotted for discrete values of $k$. Red circles correspond to a negative value, blue a positive value, and gray a value of zero, with black interpolating lines shown as a guide. (a) Resonant coupling is characterized by a relatively slow fall off in magnitude of $C_k^-$ for increasing $k$, asymptotically approaching the value $g/\sqrt{\pi\ln(k)}$ for very large $k$ and vanishing for $k\rightarrow\infty$. (b) In contrast, $|C_2^-|$ falls off very rapidly for dispersive coupling, allowing for truncation of the sum over $k$-body interactions at a small, finite value of $k$ according to the accuracy desired. The three lines show this trend for distinct values of $\lambda$, with smaller $\lambda$ displaying a faster fall off. (c) The region of parameter space for which $|C_k^-/\hbar\Gamma|>0.1$, with each colored region corresponding to a particular value of $k$. As seen explicitly for $C_2^-$ and $C_3^-$ in Fig. \ref{fig:f3}, each coefficient follows a similar trend, with areas of highest (lowest) magntitude coinciding with $\lambda\gg 1$ ($\lambda\ll 1$). For increasing $k$, the subset of parameter space in which the threshold $|C_k^-/\hbar\Gamma|>0.1$ is met tightens, with each region corresponding to order $k$ encompassing the smaller region corresponding to $k+1$. The maximal value $k=20$ was chosen for simplicity, with higher values of $k$ continuing the same trend. (d) The evolution of $|C_k^-/\hbar\Gamma|$ from resonant to dispersive coupling, shown as a function of $\Delta/\Gamma$ for constant coupling strength tuned to the onset of strong coupling $g/\Gamma = 1$ (white dashed line in panel (c)). Red and blue lines display the alternating sign of $C_k^-$. All large $k$ coefficients experience a similar, rapid fall-off as $\Delta$ is increased from the resonant case. In comparison, smaller $k$ coefficients decrease in magnitude more slowly. A ``fan out'' effect is observed as the detuning trends toward $\Delta = 10 g$, corresponding to the onset of dispersive coupling (i.e., $\lambda=0.1$). In contrast, all coefficients take on values comparable in magnitude for $\Delta=0$.}
\end{figure*}

Panels (a) and (b) of Fig. $\ref{fig:f4}$ contrast the behavior of $C_k^-$ for resonant (Eq. (\ref{eq:Ckm_res})) and dispersive (Eq. (\ref{eq:Ckm_dis})) coupling. As stated previously, the sign of the $C_k^-$ alternates in $k$ independent of parameter regime, indicated by the color of the markers. In particular, blue (red) markers represent coefficients which are positive (negative) for $\Delta\geq 0^+$, with signs inverted for $\Delta\leq0^-$. For dispersive coupling, the relative strength of $|C_k^-/\hbar\Gamma|$ falls off rapidly due to the $\lambda^{2k-1}$ dependence in Eq. (\ref{eq:Ckm_dis}) and, as a result, the sum in Eq. (\ref{eq:Hfinal}) may be truncated at some cutoff order $k_{\textrm{max}}$ dependent upon the coupling strength $g$, dispersive parameter $\lambda$, and the desired accuracy. For $k_{\textrm{max}}=2$, Eq. (\ref{eq:Hfinal}) becomes an approximate analog to the on-site portion of the Bose-Hubbard Hamiltonian in Eq. (\ref{eq:bosehubbard}). In contrast, perfectly resonant coupling is characterized by many-body coefficients $C_k^-$ which fall off slowly in $k$ and, consequently, the sum in Eq. (\ref{eq:Hfinal}) cannot be truncated unless only a finite number of excitations $n$ are considered.

For all $k$, the trend followed by the coefficients $C_k^-$ as a function of $g$ and $\Delta$ resembles that of $C_2^-$ and $C_3^-$ shown in Fig. {\ref{fig:f3}}, differing only in the rapidity with which the magnitude of $C_k^-$ falls off as $\Delta$ trends away from zero. Fig. \ref{fig:f4}c illustrates the relative magnitude of various coefficients $C_k^-$ across all parameter space, with colored sections corresponding to regions where $|C_k^-/\hbar\Gamma|\geq0.1$. Note that this threshold is somewhat arbitrary and therefore should not be taken as an exact measure of the importance of each term, as this is dependent upon the particular system and context under study. Still, the relative importance of higher order $k$-body interactions is clearly evident both for perfect resonant coupling ($\Delta=0$) and for near-resonant coupling ($\lambda\gg 1$). This is further illustrated by Fig. \ref{fig:f4} showing the magnitude of the many-body coefficients $C_k^-$ (relative to $\hbar\Gamma$) at the onset of strong coupling, $g/\Gamma = 1$, indicated by a white dotted line in Fig. \ref{fig:f4}c. Similar to panels (a) and (b), blue (red) lines indicate a positive (negative) value of $C_k^-$ for $\Delta\geq0^+$, with signs reverse for $\Delta\leq0^-$.

Finally, we note that the presented many-body form of the Jaynes-Cummings Hamiltonian must become approximately linear in the limit $n\rightarrow \infty$, inhibiting photon blockade entirely. This behavior of the Jaynes-Cummings Hamiltonian is well-known \cite{Raftery2014, Carmichael2015, greentree2006quantum} and can most easily be seen by analyzing the difference $E_{n+1,\pm}-E_{n,\pm}$ (see Eq. (\ref{eq:JCeigenval})) in the large $n$ limit. In the form Eq. (\ref{eq:Hfinal}), however, this limiting behavior is not at all obvious, particularly for resonant coupling, as the contributions of the individual, normally ordered $k$-body products $(\widetilde{a}^\dagger)^k(\widetilde{a}^\dagger)^k/k!$ return the binomial coefficient $\binom{n}{k}$ when acted on a dressed Fock state and therefore diverge for $n\to\infty$. Despite this, Eq. (\ref{eq:sumkcheck}) shows that the $k$-body interactions sum together to produce the correct eigenvalues and, as a result,
\begin{equation}
    \lim_{n \to \infty}\sum_{k=0}^{\infty} \frac{1}{k!}C_k^\pm(\widetilde{a}^\dagger)^k(\widetilde{a})^k(\ket{n+1,\pm}-\ket{n,\pm}) = 0.
\end{equation}
More qualitatively, this behavior is understood as a consequence of the alternating sign of the many-body coefficients $C_k^\pm$, causing all odd $k$-body interactions to counteract those with even $k$. As a result, the individual nonlinear interactions together conspire to give a purely linear spectrum, in alignment with the known behavior of the Jaynes-Cummings ladder in the large $n$ limit.

\section{Extension to a two site Jaynes-Cummings-Hubbard system and analysis of its quantum phases}\label{sec:2JCH}
The results of the previous section hint at a similarity between the on-site portion of the Bose-Hubbard and Jaynes-Cummings Hamiltonians -- the former containing bosonic two-body interactions, and the latter $k$-body interactions up to some order $k_{\textrm{max}}$ dependent upon the ratio $\lambda=g/\Delta$ and maximum number of excitations considered. Because only a single site was under study, the most interesting aspects of the Bose-Hubbard model, e.g., the superfluid to insulating quantum phase transition, were not discussed. The purpose of the present section is to revisit this comparison for the simplest extension possible: a two-site system. We note that qualitative and quantitative analogies between the Bose-Hubbard and Jaynes-Cummings Hubbard (JCH) models are numerous in the literature \cite{greentree2006quantum,hartmann2006strongly,hartmann2007strong,hartmann2008quantum,leib2010bose,koch2009superfluid,schmidt2009strong,Angelakis2007,carusotto2009fermionized,hohenadler2011dynamical,PhysRevLett.99.186401,PhysRevA.77.033801,PhysRevA.77.053819,hartmann2016quantum,PhysRevLett.111.160501,mering2009analytic,nietner2012ginzburg,bujnowski2014supersolid,hayward2012fractional} and, as such, we refer to these other works for a mathematically rigorous analysis of the quantum phase transition admitted by the JCH model for both a finite \cite{Angelakis2007,PhysRevA.77.053819} and infinite \cite{greentree2006quantum,koch2009superfluid,schmidt2009strong,PhysRevLett.100.216401} number of sites. Here, our aim is to illustrate the unique and complementary perspective afforded by the many-body form of the Jaynes-Cummings Hamiltonian presented in Eq. (\ref{eq:Hfinal}). Furthermore, we hope that the analysis and discussion contained herein can provide guidance for analog quantum simulators which aim to simulate many-body bosonic Hamiltonians using Jaynes-Cummings nonlinearities.

\subsection{Comparison between the Jaynes-Cummings-Hubbard and Bose-Hubbard models}\label{subsec:JCH_BH_comp}
The two-site Bose-Hubbard model is given by
\begin{equation}
    \begin{split}
        &H_{\textrm{BH}} = \sum_{i=1,2}H_{\textrm{BH,on-site}}^{(i)} + H_{\textrm{BH,hop}} \\
        &H_{\textrm{BH,on-site}}^{(i)} = -\mu b_i^\dagger b_i +  \frac{U}{2}b_i^\dagger b_i^\dagger b_i b_i \\
        &H_{\textrm{BH,hop}} = J(b_1^\dagger b_2 + b_1 b_2^\dagger).
    \end{split}
    \label{eq:twositeBH}
\end{equation}
Here, $b$ and $b^\dagger$ are bosonic annihilation and creation operators, $J$ the hopping rate between the two sites labeled $i=1,2$, and $U$ and $\mu$ are the on-site interaction strength and chemical potential, here assumed to be identical for both sites for simplicity. It is well known that this Hamiltonian admits a quantum phase transition facilitated by tuning the ratio $J/U$ at zero temperature \cite{PhysRevLett.81.3108,greiner2002quantum}. For $J\gg U$, the system is said to be in a superfluid phase, characterized by a large variance in single site particle number and a delocalized many-body ground state of the form
\begin{equation}
    \ket{\Psi_{\textrm{SF}}} \propto \left(\sum_{i=1,2}(\mp)^i b_i^\dagger \right)^n\ket{0},
    \label{eq:sfstate}
\end{equation}
where $n$ is the total number of particles in the ground state (fixed through choice of $\mu$) and the upper (lower) sign corresponds to $J>0$ ($J<0$). For simplicity, the convention $J>0$ will be assumed for the remainder of this manuscript. In the opposite limit $J\ll U$, the repulsive interaction dominates site-to-site tunneling and single site particle number fluctuations are suppressed as a result. Consequently, the ground state becomes the localized Mott-insulating state,
\begin{equation}
    \ket{\Psi_{\textrm{MI}}} \propto \prod_{i=1,2} \left(b_i^\dagger \right)^{n/2}\ket{0}.
\label{eq:mottstate}
\end{equation}

Similar to Eq. (\ref{eq:twositeBH}), the two-site JCH model may be written as
\begin{equation}
    \begin{split}
        &H_{\textrm{JCH}} = \sum_{i=1,2} H_{\textrm{JC}}^{(i)} + H_{\textrm{hop}}\\
        &H_{\textrm{JC}}^{(i)} = \hbar\omega_c a_i^\dagger a_i + \frac{1}{2}\hbar\omega_a\sigma^z_i  + \hbar g_i(a_i^\dagger \sigma_i^- + a_i \sigma_i^-)\\
        &H_{\textrm{hop}}=J(a_1^\dagger a_2 + a_1 a_2^\dagger),
    \end{split}
    \label{eq:twositeJCH}
\end{equation}
where on-site parameters $\omega_c$, $\omega_a$, and $g$ have been taken to be identical for both sites for simplicity. The parallel structure of Eq. (\ref{eq:twositeBH}) and Eq. (\ref{eq:twositeJCH}) underscores an obvious connection between the Bose-Hubbard and JCH models: both contain identical bosonic tunneling terms and similar on-site interactions, with the only distinguishing features appearing as the source of nonlinearity in $H_{\textrm{BH,on-site}}^i$ and $H_{\textrm{JC}}^i$, the former naturally including bosonic two-body terms, and the latter comprising an additional degree of freedom in the form of a TLS which ultimately mediates effective photon-photon interactions. Still, qualitative comparison between the two is merited and previous works have shown the JCH model to admit a superfluid-to-insulator quantum phase transition similar to that of the Bose-Hubbard model \cite{greentree2006quantum,hartmann2006strongly,hartmann2008quantum,koch2009superfluid,Angelakis2007,Noh2016,PhysRevLett.99.186401,PhysRevA.77.033801,PhysRevA.77.053819}, albeit with some key differences. For one, while the quantum phases of the Bose-Hubbard model are realized at opposing limits of the ratio $J/U$, the JCH model involves three distinct tunable parameters ($J$, $\Delta$, $g$) and, consequently, multiple pathways exist for tuning across a phase transition \cite{greentree2006quantum,Angelakis2007,koch2009superfluid,schmidt2009strong}. In addition, the very nature of the interacting bosonic excitations are themselves dependent upon the parameter regime, leading to a photonic-to-polaritonic transition which accompanies the superfluid-to-insulator transition in the JCH model \cite{Noh2016,Angelakis2007}, behavior which is absent in the Bose-Hubbard case. Finally, as noted previously, the JCH Hamiltonian becomes approximately linear in the limit of large $n$, while the Bose-Hubbard model maintains nonlinearity for all $n$. These distinguishing features have received qualitative recognition in the literature, yet have not been formally analyzed in the context of dressed operators where bosonic many-body interactions are brought to the forefront, as in Eq. (\ref{eq:Hfinal}). The findings of the previous section therefore compel a closer reexamination of the differences between the JCH and Bose-Hubbard models, and consequences thereof, using the techniques of unitary transformation.

Following the procedure of Section \ref{subsec:JC_diag}, we begin analysis by transforming Eq. (\ref{eq:twositeJCH}) into the dressed polariton basis. Here, we apply the transformation operator $\mathcal{U} = e^{S_1 + S_2}$ where the generator $S_i$ is defined as
\begin{equation}
    S_{i} = -\Lambda(N_i)I_{-}^{(i)}
\end{equation}
where $I_-^{(i)}$ and $\Lambda(N_i) = \theta(N_i)/\sqrt{N_i}$ are defined exactly as before (see Eqs. (\ref{eq:Hamdefs}) and (\ref{eq:mixingangle})) with the subscript $i$ inserted where appropriate to label quantities which differ between sites. For example, here $N_i$ represents the total number operator at site $i$ alone, and the total number of excitations in the system is therefore given by $N=N_1 + N_2$. It is important to emphasize that the generators $S_1$ and $S_2$ commute and the operator $\mathcal{U}$ may therefore be rewritten as a product of unitary operators $\mathcal{U}=\mathcal{U}_1\mathcal{U}_2$ where $\mathcal{U}_1=e^{S_1}$ and $\mathcal{U}_2=e^{S_2}$. Critically, $\mathcal{U}_i$ commutes with all operators associated solely with the opposite site and, as a result, transformation of the on-site contributions to the JCH proceeds exactly as in Section \ref{subsec:JC_diag}. The two-site JCH Hamiltonian may therefore be written as
\begin{equation}
    \begin{split}
        H_{\textrm{JCH}} &= H_{\textrm{hop}} +
        \sum_{i=1,2}\hbar\omega_c \left(N_i - \frac{1}{2}\right)\\
        &+ \sum_{i=1,2}\sum_{k=0}^{\infty}\frac{1}{k!}\left[ C^+_k\widetilde{\sigma}^+_i\widetilde{\sigma}^-_i + C^-_k\widetilde{\sigma}^-_i\widetilde{\sigma}^+_i\right](\widetilde{a}_i^\dagger)^k(\widetilde{a}_i)^k,
    \end{split}
    \label{eq:JCH_trans}
\end{equation}
where notation has been maintained from the previous section such that the coefficients $C_k^\pm$ are defined by Eq. (\ref{eq:mbcoefficients}) and
\begin{equation}
    \begin{split}
        \widetilde{a}_i &= \mathcal{U}^\dagger a_i \mathcal{U} \\
        \widetilde{\sigma}^-_i &= \mathcal{U}^\dagger \sigma^-_i \mathcal{U}
    \end{split}
\end{equation}
are the transformed operators describing annihilation of the dressed excitations at site $i$. We remark that Eq. (\ref{eq:JCH_trans}) is similar in form to the effective Hamiltonian presented in Ref.~\onlinecite{Noh2016} \footnote{In particular, see Eq. (5) of the referenced paper by Noh and Angelakis}. There, on-site contributions to the JCH model are written for the case $\Delta=0$ in terms of branch-dependent polaritonic operators obeying neither bosonic nor psuedo-spin commutation relations. The form presented here is therefore unique in that the dressed operators describe the true quasiparticle excitations at each site, maintaining the appropriate commutation relations, and all many-body interactions are described without use of excitation number dependent coefficients for general $g$ and $\Delta$.

Because $H_{\textrm{hop}}$ describes an exchange of purely photonic quanta, writing this explicitly in terms of dressed operators for general system parameters yields an infinite set of terms which are not obviously expressible in a closed form. Up to first order in $\Lambda$ alone, transformation of $H_{\textrm{hop}}$ yields terms corresponding to polariton hopping $J(\widetilde{a}_1^\dagger\widetilde{a}_2+\widetilde{a}_1\widetilde{a}_2^\dagger)$, linear cross-site interactions $J \Lambda(N_1)(\widetilde{a}_2^\dagger \widetilde{\sigma}_1^- + \widetilde{a}_2 \widetilde{\sigma}_1^+) + J \Lambda(N_2)(\widetilde{a}_1^\dagger \widetilde{\sigma}_2^- + \widetilde{a}_1 \widetilde{\sigma}_2^+)$, and, in addition -- because the hopping term preserves the total number of excitations $N_1 + N_2$ but not the number of excitations at each site $N_i$ -- a number of nonlinear terms which vanish in the dispersive regime (where $\Lambda(N_i)\approx\lambda$) but become important near resonant coupling. Matters are further complicated at second order in $\Lambda$, primarily due to a cascade of additional two-site terms which do not commute with $N_i$. As a result, there is little to be gained by attempting to write $H_{\textrm{hop}}$ in terms of dressed operators for general system parameters as the physics of the site to site hopping is most apparent in the bare photonic basis, and it is advantageous to instead consider several limiting cases independently. 

\subsection{Dispersive coupling: $\lambda\ll 1$}\label{subsec:2JCHdisp}
As previously discussed in Section \ref{subsec:manybodyrep}, when projected onto the negative branch the on-site terms of the JCH model directly mirror those of Bose-Hubbard model for dispersive coupling due to a sharp drop off in the coefficients $C_k^-$ for increasing $k$. Neglecting terms second order and higher in $\lambda$, the full two-site JCH Hamiltonian may be expressed as
\begin{widetext}
\begin{equation}
    \begin{split}
        H_{\textrm{JCH}} &\approx \mathcal{P}_1^-\mathcal{P}_2^-\Bigg[\sum_{i=1,2}\left(\hbar\Omega_0^{-}\widetilde{a}_i^\dagger\widetilde{a}_i + \frac{U_{\textrm{eff}}^-}{2}\widetilde{a}_i^\dagger\widetilde{a}_i^\dagger\widetilde{a}_i\widetilde{a}_i\right)  + J(\widetilde{a}_1^\dagger\widetilde{a}_2+\widetilde{a}_1\widetilde{a}_2^\dagger) + 2E_0^-\Bigg] \\
        &+ \mathcal{P}_1^+\mathcal{P}_2^+\Bigg[\sum_{i=1,2}\left(\hbar\Omega_0^{+}\widetilde{a}_i^\dagger\widetilde{a}_i + \frac{U_{\textrm{eff}}^+}{2}\widetilde{a}_i^\dagger\widetilde{a}_i^\dagger\widetilde{a}_i\widetilde{a}_i\right)  + J(\widetilde{a}_1^\dagger\widetilde{a}_2+\widetilde{a}_1\widetilde{a}_2^\dagger) + 2E_0^+\Bigg]\\
        &+ \mathcal{P}_1^+\mathcal{P}_2^-\Bigg[\hbar\Omega_0^{+}\widetilde{a}_1^\dagger\widetilde{a}_1 + \frac{U_{\textrm{eff}}^+}{2}\widetilde{a}_1^\dagger\widetilde{a}_1^\dagger\widetilde{a}_1\widetilde{a}_1 + \hbar\Omega_0^{-}\widetilde{a}_2^\dagger\widetilde{a}_2 + \frac{U_{\textrm{eff}}^-}{2}\widetilde{a}_2^\dagger\widetilde{a}_2^\dagger\widetilde{a}_2\widetilde{a}_2 + J(\widetilde{a}_1^\dagger\widetilde{a}_2+\widetilde{a}_1\widetilde{a}_2^\dagger) + E_0^+ + E_0^-\Bigg]\\ &+\mathcal{P}_1^-\mathcal{P}_2^+\Bigg[\hbar\Omega_0^{-}\widetilde{a}_1^\dagger\widetilde{a}_1 + \frac{U_{\textrm{eff}}^-}{2}\widetilde{a}_1^\dagger\widetilde{a}_1^\dagger\widetilde{a}_1\widetilde{a}_1 + \hbar\Omega_0^{+}\widetilde{a}_2^\dagger\widetilde{a}_2 + \frac{U_{\textrm{eff}}^+}{2}\widetilde{a}_2^\dagger\widetilde{a}_2^\dagger\widetilde{a}_2\widetilde{a}_2 + J(\widetilde{a}_1^\dagger\widetilde{a}_2+\widetilde{a}_1\widetilde{a}_2^\dagger) + E_0^+ + E_0^-\Bigg]\\
        &+ J\lambda(\widetilde{a}_1^\dagger\widetilde{\sigma}_2^- + \widetilde{a}_1\widetilde{\sigma}_2^+) + J\lambda(\widetilde{a}_2^\dagger\widetilde{\sigma}_1^- + \widetilde{a}_2\widetilde{\sigma}_1^+)
    \end{split}
    \label{eq:JCH_dispersive}
\end{equation}
\end{widetext}
where $\Omega_0^{\pm}=\omega_c+C_1^\pm/\hbar$ denotes an effective resonant energy,  $U_{\textrm{eff}}^{\pm}=C_2^\pm$ an effective interaction strength, $E_0^\pm = C_0^\pm \pm \hbar\omega_c/2$ a constant energy shift, and $\mathcal{P}^\pm_i = \widetilde{\sigma}_i^\pm\widetilde{\sigma}_i^\mp$ the projector onto the positive (upper sign) or negative (lower sign) branch of the $i$th site.

In writing Eq. (\ref{eq:JCH_dispersive}), the on-site terms were cast into the dressed basis using the techniques of Section \ref{subsec:manybodyrep}, while the hopping Hamiltonian $H_{\textrm{hop}}$ was reexpressed in terms of dressed operators using the transformed form of the relations Eq. (\ref{eq:ops_approx_disp}). Thus, using the techniques presented here, we have made the analogy between the two-site JCH and Bose-Hubbard models for $\lambda\ll1$ as explicit as possible -- Eq. (\ref{eq:JCH_dispersive}) illustrates that, in the dispersive limit, the two-site JCH describes physical behavior which mirrors the Bose-Hubbard model independently within each of its four branches. Interestingly, these four branches allow for realization of either a symmetric ($\propto\mathcal{P}^\pm\mathcal{P}^\pm$) or an asymmetric ($\propto\mathcal{P}^\pm\mathcal{P}^\mp$) Bose-Hubbard type system. Unlike the single Jaynes-Cummings Hamiltonian, however, transitions between the various branches are allowed due to the cross-site boson-TLS couplings induced by transformation of $H_{\textrm{hop}}$. This effect was previously noted and analyzed in Refs.~\onlinecite{Angelakis2007} and~\onlinecite{koch2009superfluid}, there described in the context of polariton operators as an interconversion between $+$ and $-$ polariton types. Due to their scaling with $\lambda\ll1$, these terms only weakly contribute in comparison to the dressed bosonic hopping term $J(\widetilde{a}_1^\dagger\widetilde{a}_2 + \widetilde{a}_1\widetilde{a}_2^\dagger)$ for arbitrary $J$. This fact is not unsurprising as, in the dispersive regime, the dressed bosonic operators are photon-like. As a result, the purely photonic hopping term $H_{\textrm{hop}}$ is well-approximated by a photon-like dressed bosonic hopping and, consequently, the influence of the last two terms of Eq. (\ref{eq:JCH_dispersive}) may be approximated using second-order perturbation theory or, depending on the value of $\lambda$ and the accuracy desired, entirely neglected.

It is clear from a qualitative argument alone that the Hamiltonian in Eq. (\ref{eq:JCH_dispersive}) admits an insulator-to-superfluid transition analogous to that of the Bose-Hubbard model largely unaltered by the final two inter-branch terms: focusing on the branch corresponding to the projector $\mathcal{P}_1^- \mathcal{P}_2^-$ and recalling that $C_2^-/\hbar\approx 2\lambda^3 g$ in the dispersive regime, the limit $J/U_{\textrm{eff}} \ll 1$ (equivalent to $J/\hbar g\ll 2\lambda^3$ in terms of basic system parameters) yields a localized, insulator-like $n$ particle ground state identical to Eq. (\ref{eq:mottstate}). In the opposite limit $J/U_{\textrm{eff}}^-\gg 1$ (or identically, $J/\hbar g\gg 2\lambda^3$), the influence of the inter-branch terms is felt only at second order perturbation theory in $\lambda$. For $\lambda\ll 1$, the $n$ particle ground state becomes identical to the delocalized, superfluid-like, state Eq. (\ref{eq:sfstate}).

\subsection{The $n\leq2$ limit}\label{subsec:2JCH_neq2}
As evidenced in Section \ref{subsec:manybodyrep}, another useful strategy for theoretical analysis involves truncating the composite Hilbert space by restricting the total number of excitations to a finite, maximal value. This approach is particularly relevant for comparison with the Bose-Hubbard model, where the chemical potential $\mu$ naturally determines the total number of particles in the many-body ground state \cite{fetter2012quantum}. Fixing to a particular excitation number in the JCH Hamiltonian therefore facilitates a straightforward comparison. Furthermore, truncating the Hilbert space allows for closer inspection of the resonant coupling regime which, as previously discussed, is challenging to analyze for general $n$ due to the difficulty of casting $H_{\textrm{hop}}$ in terms of dressed operators for general system parameters. For simplicity, we specialize to the case of $n\leq 2$. As shown in Section \ref{subsec:manybodyrep}, this limit results in an exact analogy between the on-site terms of the Bose-Hubbard model and the negative branch of the many-body representation of the Jaynes-Cummings model for both resonant and dispersive coupling. 

It is convenient to first reexpress the two-site JCH Hamiltonian in terms of projectors onto the positive and negative branch at each site:
\begin{equation}
    H_{\textrm{JCH}} = \sum_{s_1,s_2,s_1',s_2'}\ket{s_1',s_2'}\bra{s_1',s_2'}H\ket{s_1,s_2}\bra{s_1,s_2}.
    \label{eq:Hmatrix}
\end{equation}
Here, each $s_i$ is summed over the values $+$ and $-$, the first and second entry of each bra/ket indicate the state of the TLS at the first and second sites, and the subscript ``JCH'' has been dropped from the various matrix elements for simplicity. For convenience, we define the notation $\mathcal{H}_{s_1 s_2}$ to represent to subspace spanned by the states $\{\ket{m_1, m_2, s_1, s_2}\}$, where the four indices denote, in order, the eigenvalues of $\widetilde{a}_1^\dagger\widetilde{a}_1$, $\widetilde{a}_2^\dagger\widetilde{a}_2$, $\widetilde{\sigma}_1^z$, and $\widetilde{\sigma}_2^z$. Note that here we are adopting notation for the states which is slightly modified from Section \ref{sec:JC}, as the first index no longer corresponds to the total number of excitations at the $i$th site but rather the number of quanta in the dressed bosonic mode alone. This simplification is made both to avoid confusion with the prefactors returned by dressed operators (see Table \ref{tab:ops}), but also because $N_1$ and $N_2$ are no longer independently conserved quantities and therefore their notational utility is diminished. 

Drawing upon the discussion of the Jaynes-Cummings Hamiltonian in Section \ref{subsec:manybodyrep}, it is the subspace $\mathcal{H}_{--}$ which includes the vacuum state. Consequently, our analysis will focus on the many-body physics within this subspace. As previously shown for the case of dispersive coupling, the two site JCH Hamiltonian differs from the single site Jaynes-Cummings Hamiltonian in that inter-branch transitions can occur due to nonzero off-diagonal elements of Eq. (\ref{eq:Hmatrix}) contributed by the purely photonic hopping term.  In order to simplify discussion of these matrix elements in the present formalism, we introduce the notation
\begin{equation}
    \bar{H}\equiv\bra{-,-}H\ket{-,-}
\end{equation}
to denote the block of $H_{\textrm{JCH}}$ which contributes to dynamics confined within the target subspace $\mathcal{H}_{--}$. Similarly, let
\begin{equation}
    V_{s_1s_2}\equiv\bra{s_1,s_2}H\ket{-,-}
\end{equation}
denote the set of matrix elements describing allowed transitions from $\mathcal{H}_{--}$ to its complement $\mathcal{H}_{+-}\cup\mathcal{H}_{-+}\cup\mathcal{H}_{++}$. Because $H_{\textrm{JCH}}$ is Hermitian, for every allowed transition from $\mathcal{H}_{--}$ to $\mathcal{H}_{s_1s_2}$, there exists a transition of equal probability describing the inverse process described by the matrix elements of $V_{s_1s_2}^\dagger$.

\begin{figure*}
\includegraphics[width=0.55\textwidth]{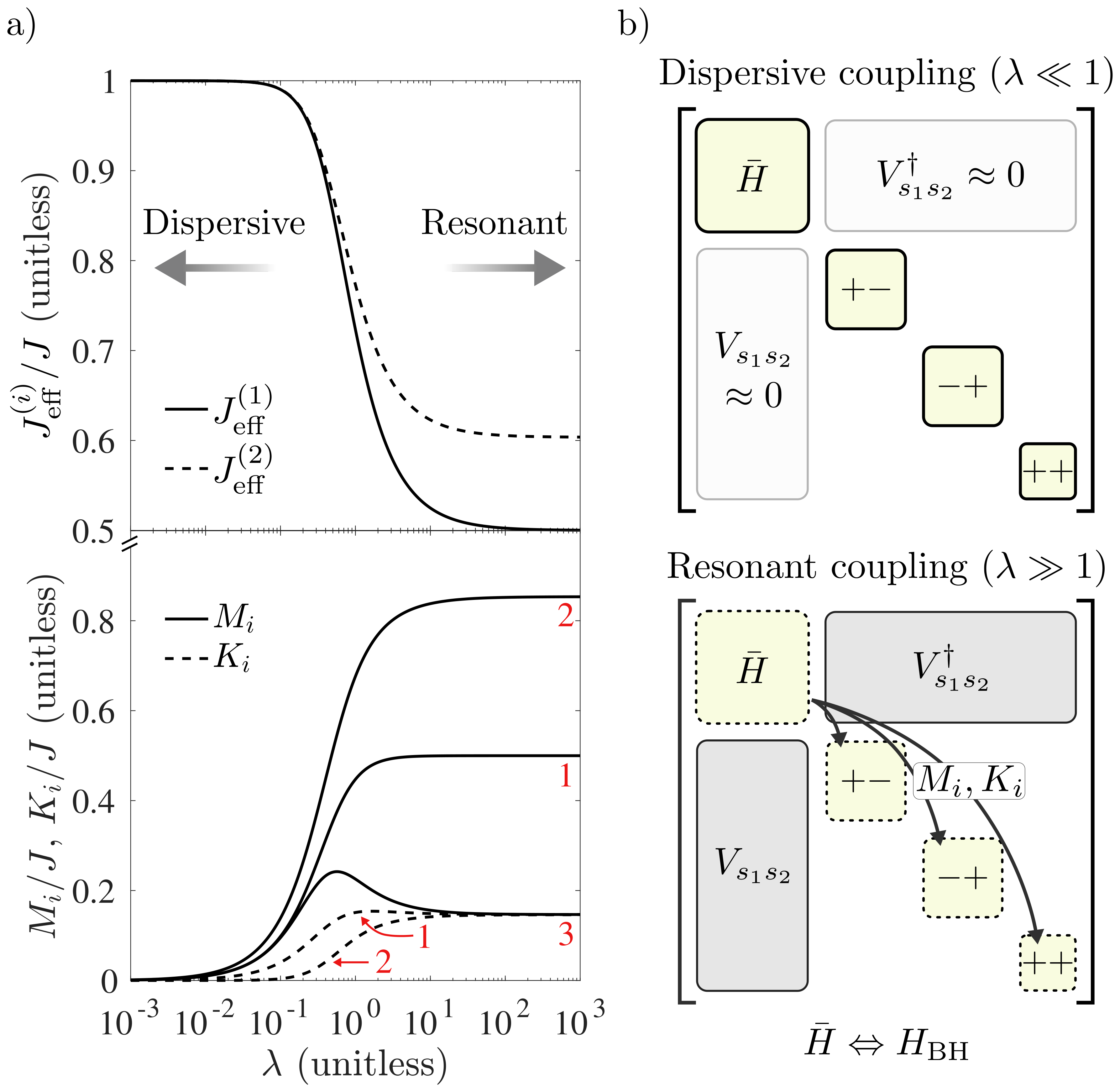}
\caption{\label{fig:f5} (a) The top panel shows the one and two excitation effective hopping strengths $J_{\textrm{eff}}^{(1)}$ and $J_{\textrm{eff}}^{(2)}$ while the bottom panel displays the value of the three unique linear ($M_1$, $M_2$, $M_3$) and nonlinear ($K_1$, $K_2$) transition amplitudes $M_1$, $M_2$, $M_3$, $K_1$ and $K_2$. All are plotted as a function of $\lambda$ and normalized to the bare photonic hopping rate $J$. Notably, the first set of parameters $J_{\textrm{eff}}^{(i)}$ describe the amplitude of \emph{intra}-branch transitions within the subspace $\mathcal{H}_{--}$, while the second set describes the amplitudes of \emph{inter}-branch transitions between the subspace $\mathcal{H}_{--}$ and its complement $\mathcal{H}_{+-}\cup\mathcal{H}_{-+}\cup\mathcal{H}_{++}$ via cross-site light-matter interactions in the dressed representation. In the dispersive limit, all inter-branch transition amplitudes become small and the effective hopping strengths tend toward $J$. In the opposite regime, $J_{\textrm{eff}}^{(1)}$ and $J_{\textrm{eff}}^{(2)}$ differ by scalar prefactors and the coefficients $M_i$ and $K_i$ become comparable to $J$, leading to appreciable dissimilarity with the two-site Bose-Hubbard model. (b) Illustration of the two-site JCH Hamiltonian $H_{\textrm{JCH}}$ in the dispersive (top panel) and resonant (bottom panel) coupling regimes. For the truncated space of two or fewer total excitations, each panel represents a $13\times13$ matrix comprising the four branches shown along each diagonal and denoted by the state of each dressed TLS. Note that in place of ``$--$'' is the $6 \times 6$ matrix $\bar{H}$, defined via the projection of the two-site JCH onto the target subspace $\mathcal{H}_{--}$. In general, the ten nonvanishing matrix elements of $V_{s_1s_2}$ yield five unique values given by the coefficients $M_i$ and $K_i$ defined in Eq. (\ref{eq:MKdefs}). In the dispersive regime, $H_{\textrm{JCH}}$ becomes approximately block diagonal in the dressed basis as $M_i/J\ll 1$ and $K_i/J \ll 1$ and inter-branch transitions become negligible. As a result, $\bar{H}$ becomes an appropriate effective Hamiltonian and analogy to the Bose-Hubbard model is realized. In the resonant regime, inter-branch transitions become important and, consequently, direct correspondence with the Bose-Hubbard model collapses.}
\end{figure*}

Constraining the full Hilbert space to two or fewer excitations, the projection of $H$ onto the subspace $\mathcal{H}_{--}$ may be expressed in the block-diagonal form
\begin{equation}
    \bar{H} = \left[
        \begin{matrix}
             \bar{H}_{n=0}    & 0                 & 0       \\
             0                & \bar{H}_{n=1}     & 0       \\
             0                & 0                 & \bar{H}_{n=2}      \\
             
        \end{matrix}
    \right]
    \label{eq:Hbar}
\end{equation}
where $\bar{H}_{n}$ is a square matrix of dimension $2^n$ corresponding to the subspace of $n$ excitations, containing diagonal and off-diagonal entries given by the on-site and hopping terms of Eq. (\ref{eq:JCH_trans}), respectively. Because the JCH conserves the total number of excitations, transitions between states of different total particle number $n$ are not allowed, hence the block-diagonal form of Eq. (\ref{eq:Hbar}). Discarding the vacuum energy $\bar{H}_{n=0}=2C_0^- -\hbar\omega_c$, the single and double excitation blocks of $\bar{H}$ may be written in the form
\begin{equation}
    \begin{gathered}
            \bar{H}_{n=1} = \left[\begin{matrix}
                 \hbar\Omega_0 & J_{\textrm{eff}}^{(1)} \\
                 J_{\textrm{eff}}^{(1)} & \hbar\Omega_0
            \end{matrix}\right] \\
            \bar{H}_{n=2} = \left[\begin{matrix}
                 2\hbar\Omega_0 + U_{\textrm{eff}} & 0 & \sqrt{2}J_{\textrm{eff}}^{(2)} \\
                 0 & 2\hbar\Omega_0 + U_{\textrm{eff}} & \sqrt{2}J_{\textrm{eff}}^{(2)} \\
                 \sqrt{2}J_{\textrm{eff}}^{(2)} & \sqrt{2}J_{\textrm{eff}}^{(2)} & 2\hbar\Omega_0
            \end{matrix}\right].
    \end{gathered}
    \label{eq:JCHeff_nleq2_proj}
\end{equation}
Here, the effective on-site resonant energy and interaction strength are $\Omega_0=\omega_c+C_1^-/\hbar$ and $U_{\textrm{eff}}=C_2^-$, where negative sign superscripts have been removed from $\Omega_0$ and $U_{\textrm{eff}}$ relative to Eq. (\ref{eq:JCH_dispersive}) for simplicity. In addition, the effective hopping strengths are defined by
\begin{equation}
    \begin{split}
        J_{\textrm{eff}}^{(1)} &= J\cos^2\theta{(1)} \\
        J_{\textrm{eff}}^{(2)} &= J\cos\theta{(1)}\big[\cos\theta{(1)}\cos\theta{(2)}+\sin\theta{(1)}\sin\theta{(2)/\sqrt{2}}\big],
    \end{split}
\end{equation}
where $\theta(N)$ is the mixing angle previously defined in Eq. (\ref{eq:mixingangle}) and the vector space is ordered as $\left\{ \ket{10},\ket{01}\right\}$ for $n=1$ and $\left\{ \ket{20},\ket{02},\ket{11}\right\}$ for $n=2$. We emphasize again that Eq. (\ref{eq:JCHeff_nleq2_proj}) does not fully describe the dynamics of the two-site JCH, even for the limit $n\leq2$, due to possible transitions to and from the target subspace $\mathcal{H}_{--}$ described by $V_{s_1s_2}$. Still, it is useful to first examine the similarities between Eq. (\ref{eq:JCHeff_nleq2_proj}) and the Bose-Hubbard model in isolation. Making as explicit a comparison as possible, projecting the Bose-Hubbard Hamiltonian onto the subspace $n\leq2$ leads to
\begin{equation}
    \begin{gathered}
            H_{\textrm{BH},n=1} = \left[\begin{matrix}
                 -\mu & J \\
                 J & -\mu
            \end{matrix}\right] \\
            H_{\textrm{BH},n=2} = \left[\begin{matrix}
                 -2\mu + U & 0 & \sqrt{2}J \\
                 0 & -2\mu + U & \sqrt{2}J \\
                 \sqrt{2}J & \sqrt{2}J & -2\mu
            \end{matrix}\right].
    \end{gathered}
    \label{eq:BHmatrixform}
\end{equation}
Though nearly identical in form to Eq. (\ref{eq:JCHeff_nleq2_proj}), a few key differences must be highlighted. First, the one and two excitation manifolds of the two-site JCH are characterized by different tunneling strengths $J_{\textrm{eff}}^{(1)}$ and $J_{\textrm{eff}}^{(2)}$ which identically approach $J$ in the dispersive limit, but plateau to different values for the resonant case (see Fig. \ref{fig:f5}a, top panel). Second, as already emphasized, underlying the two-site JCH is a larger range of independently tunable parameters ($\omega_c$, $\Delta$, $g$, $J$) compared to the two-site Bose-Hubbard model which is characterized by $\mu$, $U$ and $J$ alone. At the level of the effective parameters in Eq. (\ref{eq:JCHeff_nleq2_proj}), however, it is important to be mindful -- particularly for the purpose of quantum simulation of Bose-Hubbard models with JCH systems -- that tuning $U_{\textrm{eff}}$ while holding $J_{\textrm{eff}}$ constant, for example, requires the explicit understanding of how these effective parameters depend upon those in the fundamental parameters (e.g., $\Delta$, $g$) as demonstrated here. Likewise, changing $g$ while holding $\Delta$ constant can impact not only the effective repulsion strength $U_{\textrm{eff}}$, as expected, but also the effective two excitation tunneling strength $J_{\textrm{eff}}^{(2)}$. Thus, the analytic forms for these parameters is not of just theoretical, but also experimental interest. 

\begin{figure*}
\includegraphics[width=1\textwidth]{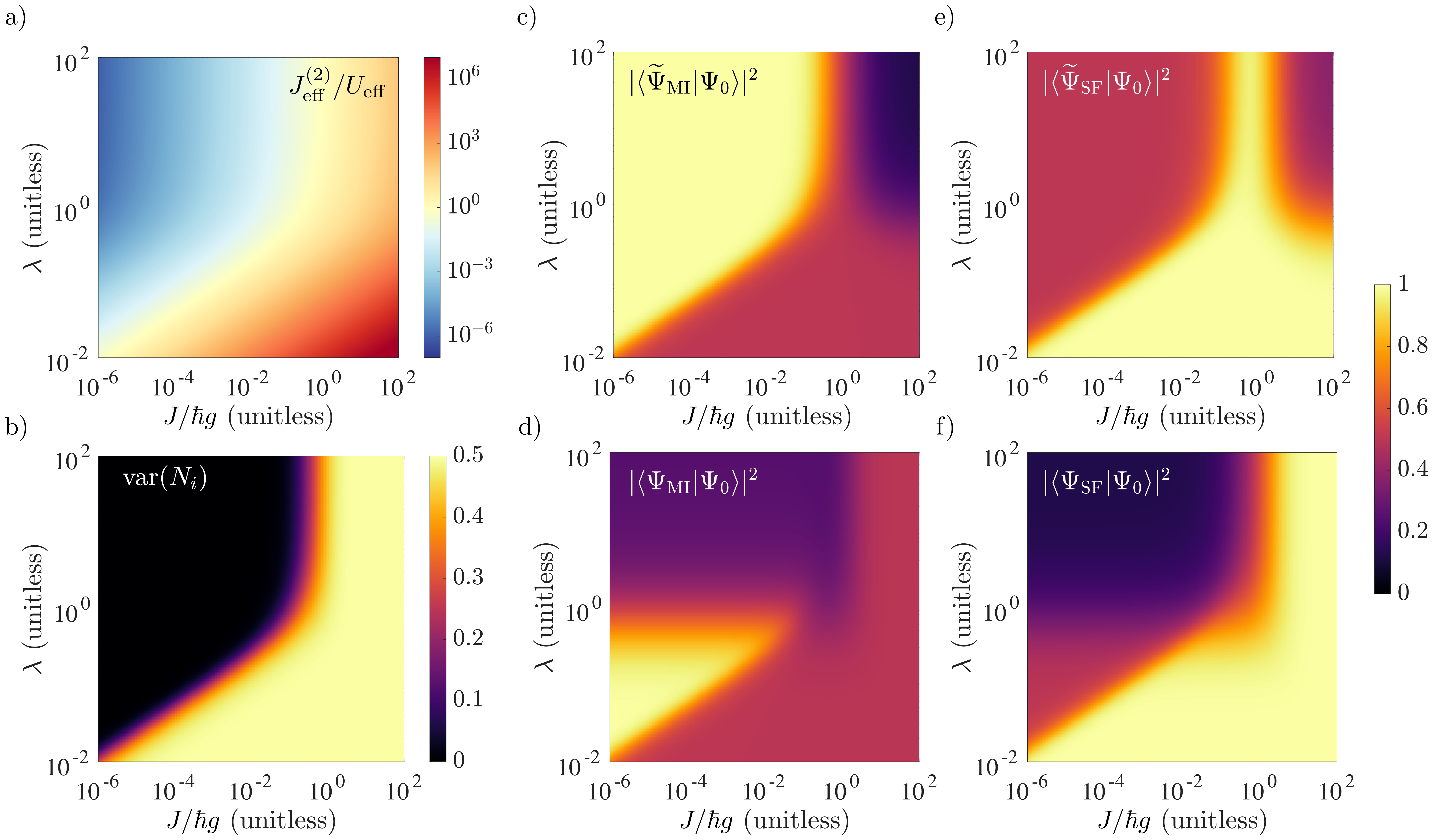}
\caption{\label{fig:f6} (a) Ratio of the analytically derived effective two excitation hopping rate $J_{\textrm{eff}}^{(2)}$ and dressed boson-boson interaction strength $U_{\textrm{eff}}$ as a function of $\lambda$ and $J/\hbar g$. (b) Variance of the total number of excitations at site $i$ as a function of $\lambda$ and $J/\hbar g$. Due to symmetry there is no distinction in the variance at site one or two. This quantity serves as a useful order parameter for finite lattice systems, with $\textrm{var}(N_i)$ vanishing for a Mott-insulating state but taking a finite value for a superfluid, reaching $\textrm{var}(N_i)=0.5$ for an ideal $n=2$ superfluid state. Clearly, panel (b) suggests the possibility for both an insulating and superfluid phase in either the dispersive or resonant regimes independently. Comparing to panel (a), regions of $J_{\textrm{eff}}^{(2)}/U_{\textrm{eff}}\ll 1$ and $J_{\textrm{eff}}^{(2)}/U_{\textrm{eff}}\gg 1$ correlate near perfectly with regions of vanishing and nonvanishing variance, respectively, with $J_{\textrm{eff}}^{(2)}/U_{\textrm{eff}}\sim1$ demarcating the boundary. (c--f) Overlap of the two-particle ground state $\ket{\Psi_0}$ with the idealized (c) dressed Mott-insulating, (d) photonic Mott-insulating, (e) dressed superfluid and (f) photonic superfluid states, defined in Eq. (\ref{eq:gndstatecmps}), as a function of $\lambda$ and $J/\hbar g$. Depending on the value of $\lambda$, the dressed states take on a either a photonic ($\lambda\ll 1$) or ($\lambda\gg 1$) polaritonic character, where this former limit is responsible for the noticeable agreement between top and bottom panels for $\lambda\ll 1$. In the dispersive regime, tuning from small to large values of $J/\hbar g$ facilitates a transition from a photonic Mott-insulating phase to a photonic superfluid phase. In the resonant coupling regime, tuning $J/\hbar g$ results in three distinct phases, with a polaritonic Mott-insulating state occuring for $J_{\textrm{eff}}\ll1$, a polaritonic superfluid for $J_{\textrm{eff}}\sim1$, and a photonic superfluid $J_{\textrm{eff}}\gg1$.}
\end{figure*}

To better illustrate the nontrivial relationship between the effective parameters ($U_{\textrm{eff}}$ and $J^{(2)}_{\textrm{eff}}$) and their base parameter counterparts ($g$ and $J$), Fig. \ref{fig:f6}a displays the ratio  $J_{\textrm{eff}}^{(2)}/U_{\textrm{eff}}$ as a function of $\lambda$ and $J/\hbar g$. In computing these values, the cavity resonant frequency and light-matter coupling strength were fixed at $\omega_c/\Gamma=10^{3}$ and $g/\Gamma = 1$ while $\Delta$ and $J$ were allowed to vary. Notably, the limits $J_{\textrm{eff}}^{(2)}/U_{\textrm{eff}}\ll 1$ and $J_{\textrm{eff}}^{(2)}/U_{\textrm{eff}}\gg 1$, relevant for realization of Mott-insulating-like and superfluid-like phases, are reached not just through choice of $J/\hbar g$ but also $\lambda$. Fig. \ref{fig:f5}b therefore serves to illustrate the complexity in navigating the comparably larger parameter space of the JCH model for realization of behavior analogous to the Bose-Hubbard model, while also serving as a useful guide for achieving a particular effective parameter regime of interest.

In isolation, Eqs. (\ref{eq:JCHeff_nleq2_proj}) and (\ref{eq:BHmatrixform}) define Hamiltonians closely mirroring one another and thus describe analogous physical behavior. However, a more honest comparison must take into account the matrix elements of $V_{s_1s_2}$. In total, there are ten unique transitions (twenty including the reverse processes described by $V_{s_1s_2}^\dagger$), all of which may be divided into two categories: linear cross-site bosonic-TLS couplings and more complicated nonlinear interactions involving both on-site and cross-site exchange of quanta. Symmetry of the two sites dictates that each transition is accompanied by a parity reversed pair. All twenty allowed transitions may therefore be summarized by the outcoupling coefficients
\begin{equation}
    \begin{split}
        M_1 &\equiv \bra{0,0}V_{+-}\ket{0,1}=\bra{0,0}V_{-+}\ket{1,0} \\
        M_2 &\equiv \bra{0,1}V_{+-}\ket{0,2}=\bra{1,0}V_{-+}\ket{2,0} \\
        M_3 &\equiv \bra{1,0}V_{+-}\ket{1,1}=\bra{0,1}V_{-+}\ket{1,1} \\
        K_1 &\equiv \bra{0,1}V_{+-}\ket{2,0}=\bra{1,0}V_{-+}\ket{0,2} \\
        K_2 &\equiv \bra{0,0}V_{++}\ket{2,0}=\bra{0,0}V_{++}\ket{0,2} \\
    \end{split}
    \label{eq:MKdefs}
\end{equation}
and their Hermitian conjugates, where $M_i$ denotes a linear cross-site interaction (i.e., exchange of a single quantum) and $K_i$ labels a nonlinear process (i.e., exchange of multiple quanta). All are similar in form to the effective hopping strength $J_{\textrm{eff}}^{(i)}$ -- proportional to $J$ but otherwise dependent only on the mixing angle $\theta$ or, equivalently, $\lambda=g/\Delta$, via products of trigonometric functions. For explicit analytic forms of each coefficient, see Appendix \ref{sec:App2}.

As shown in the bottom panel of Fig. \ref{fig:f5}a, the magnitude of the coefficients $M_i$ and $K_i$ depend drastically on $\lambda$ and, as a result, the transition rate out of the subspace $\mathcal{H}_{--}$ differs between the dispersive and resonant coupling regimes. In the former case, the various outcouplings may be approximated to lowest order as
\begin{equation}
    \begin{split}
        M_1 &\approx M_3 \approx \lambda J\\
        M_2 &\approx \sqrt{2}\lambda J, \\
    \end{split}
    \label{eq:outcoupling_dispersive}
\end{equation}
where the nonlinear transition amplitudes $K_1 \approx \mathcal{O}(\lambda^3)$ and $K_2 \approx \mathcal{O}(\lambda^4)$ are comparatively small and may therefore be neglected. This result is in agreement with the more general Hamiltonian of Eq. (\ref{eq:JCH_dispersive}). As previously established, tunneling within the subspace $\mathcal{H}_{--}$ clearly dominates outcoupling for $\lambda\ll1$ and, to first approximation, Eq. (\ref{eq:Hbar}) serves as an appropriate effective Hamiltonian without consideration of outcouplings. It is interesting to note, however, that the outcouplings which contribute most meaningfully -- namely, $M_1$, $M_2$ and $M_3$ -- all resemble single excitation losses from the perspective of the dressed bosons. This suggests the possibility for a non-perturbative treatment via projective methods, ultimately leading to a repackaging at the level of effective dissipation rates and energy shifts which renormalize the matrix elements of Eq. (\ref{eq:Hbar}) \cite{cohen1998atom}. This approach would not qualitatively alter the parallel structure with the Bose-Hubbard model, however, so we leave the described strategy as a possible future avenue for analyses where quantitative agreement is desired.

In contrast with the dispersive case, resonant coupling is characterized by nonvanishing linear and nonlinear transition amplitudes which plateau to values of order $J$. In the limit $U_{\textrm{eff}}\gg J$, these contributions are unimportant as ${M_i,K_i}<J$ for all $\lambda$ and the system is therefore dominated by on-site interactions, leading to a Mott-insulating, two particle ground state comprising polaritonic excitations. In the opposite limit $U_{\textrm{eff}}\ll J$, however, Eq. (\ref{eq:JCHeff_nleq2_proj}) fails to capture the entirety of the dynamics due to the importance of inter-branch transitions, as illustrated in Fig. \ref{fig:f5}c.

\subsection{The $n=2$ ground state: quantum phases in the dispersive and resonant coupling regimes}\label{subsec:2JCHphases}

To better understand the turn on of these inter-branch transitions  and their impact on the quantum phase transition admitted by the two-site JCH, we numerically compute the two-particle ground state as a function of $J/\hbar g$ and $\lambda$ taking into account both intra- and inter-branch dynamics of Eq. (\ref{eq:JCH_trans}). In general, this ground state is a superposition of the eight possible two excitation states \{$\ket{2,0,-,-}$, $\ket{0,2,-,-}$, $\ket{1,1,-,-}$, $\ket{1,0,+,-}$, $\ket{0,1,+,-}$, $\ket{1,0,-,+}$, $\ket{0,1,-,+}$,  $\ket{0,0,+,+}$\}, gaining contributions not only from the target subspace $\mathcal{H}_{--}$ but also its complement. Similar to Fig. \ref{fig:f5}b, all calculations were carried out for the fixed values $\omega_c/\Gamma = 10^3$ and $g/\Gamma = 1$, allowing $\Delta$ and $J$ to independently vary. Following Ref.~\onlinecite{Angelakis2007}, we use the variance in particle number at the $i$th site, $\textrm{var}(N_i)$, as an order parameter. The computed variance for the two particle ground state is shown in Fig. \ref{fig:f6}b as a function of $J/\hbar g$ and $\lambda$.

Comparing Figs. \ref{fig:f6}a and \ref{fig:f6}b, it is clear that the variance tracks the value of $J_{\textrm{eff}}^{(2)}/U_{\textrm{eff}}$, as would be expected in the Bose-Hubbard model, with regions of vanishing variance (i.e., a Mott-insulator) corresponding to $J_{\textrm{eff}}^{(2)}/U_{\textrm{eff}}\ll 1$ and regions which plateau to $\textrm{var}(N_i)=0.5$ (i.e., a superfluid) where $J_{\textrm{eff}}^{(2)}/U_{\textrm{eff}}\gg 1$. In the dispersive regime, the onset of the superfluid-like phase occurs at increasingly smaller values of $J/\hbar g$ as $\lambda$ is decreased. This phenomenon is easily understood through appeal to the analytic correspondence $U_{\textrm{eff}}=C_2^-$ and reference to previously derived results. In particular, in Section \ref{subsec:manybodyrep} it was shown that $C_2^-/\hbar\approx2\lambda^3 g$ in the dispersive regime. Then simultaneously maintaining a constant photon-photon interaction strength ($C_2^-$) while increasing $\Delta$ (i.e., decreasing $\lambda$) requires a relative increase in $g$, pushing the regime where photon-photon interactions dominate over photonic hopping toward smaller values of $J/\hbar g$ as as the system moves further into dispersive coupling. Oppositely, on resonance, it was found that $C_2^-/\hbar= (2-\sqrt{2})g$. Then $U_{\textrm{eff}}$ is equivalent to $g$ up to some scalar prefactor and the phase transition will occur at roughly the same value of $J/\hbar g$ for all $\lambda\gg 1$.

It is important to reemphasize that $N_i$ commutes with the unitary transformation operator $\mathcal{U}=e^{S_1+S_2}$, and thus Fig. \ref{fig:f6}b may equally well be interpreted as the variance in the total number of bare or dressed photonic and atomic excitations. Given this, it is notable that the ``phase boundary'' is qualitatively demarcated by the $J_{\textrm{eff}}^{(2)}/U_{\textrm{eff}}=1$ line, entirely dependent on effective parameters appearing in the dressed basis. This agreement therefore indicates not only that the effective parameters $J_{\textrm{eff}}^{(2)}$ and $U_{\textrm{eff}}$ analytically derived here are the appropriate JCH model counterparts of the Bose-Hubbard parameters $J$ and $U$, but also that the dressed operator basis provides the most approrpriate representation for understanding the many-body phenomena of the JCH.

Because the order parameter $\textrm{var}(N_i)$ does not distinguish between excitations which are photonic in nature and those which are polaritonic, Fig. \ref{fig:f6}b provides little insight into the physical makeup of the excitations composing the two-particle ground state. For instance, the black, $\textrm{var}(N_i)=0$ region of Fig. \ref{fig:f6}b clearly suggests that the system is in an insulating state, characterized by a constant number of quanta at each site. It does not, however, provide any information about whether these excitations are fundamentally photonic or polaritonic. In order to gain a deeper understanding of the ground state, we compute its squared overlap with the four distinct states
\begin{subequations}
\begin{align}
    &\ket{\widetilde{\Psi}_{\textrm{MI}}} = \widetilde{a}_1^\dagger \widetilde{a}_2^\dagger \ket{0}\\
    &\ket{\Psi_{\textrm{MI}}} = a_1^\dagger a_2^\dagger \ket{0} \\
    &\ket{\widetilde{\Psi}_{\textrm{SF}}} = \frac{1}{2\sqrt{2}}(\widetilde{a}_1^\dagger - \widetilde{a}_2^\dagger)^2 \ket{0}\\
    &\ket{\Psi_{\textrm{SF}}} = \frac{1}{2\sqrt{2}}(a_1^\dagger - a_2^\dagger)^2 \ket{0}
    \end{align}
    \label{eq:gndstatecmps}
\end{subequations}
which denote Mott-insulating (a,b) and superfluid-like (c,d) states in both the bare and dressed dressed excitation bases via action of the appropriate creation operators on the vacuum state $\ket{0} = \ket{0,0,g,g} = \ket{0,0,-,-}$. 

Figs. \ref{fig:f6}c and \ref{fig:f6}e show the squared projection of the computed ground state onto the dressed Mott-insulating and superfluid states, while Figs. \ref{fig:f6}d and \ref{fig:f6}f show the corresponding projections onto their bare photonic counterparts. Focusing first on the dispersive regime (i.e., roughly the bottom third of each plot), comparison of the upper and lower panels agrees with theoretical intuition -- for $\lambda\ll1$, the dressed basis is a merely perturbed version of the bare basis due to the weak light-matter mode mixing and, as a result, there is little distinction between the bare and dressed photons. Using Fig. \ref{fig:f6}a as a visual guide, regions where $J_{\textrm{eff}}/U_{\textrm{eff}}\ll 1$ correspond to near unity overlap with the photonic Mott-insulating state $\ket{\Psi_{\textrm{MI}}}$ while regions of $J_{\textrm{eff}}/U_{\textrm{eff}}\gg 1$ perfectly conform to the photonic superfluid state $\ket{\Psi_{\textrm{SF}}}$. The phase boundary occurs roughly at $J_{\textrm{eff}}/U_{\textrm{eff}}\approx 1$, further establishing the utility of the analytic mapping between basic system parameters and the effective Bose-Hubbard like parameters presented here. Thus, the quantum phase transition as $J/\hbar g$ is tuned for constant $\lambda\ll 1$ behaves exactly as predicted in Sec. \ref{subsec:2JCH_neq2}.

In the resonant coupling regime (roughly the top third of each plot), the physical character of bare and dressed excitations fundamentally differ as $\theta\approx \pi/4$ and the operators $\widetilde{a}_i^\dagger$ and $\widetilde{a}_i$ therefore describe creation and annihilation of polaritons. This divergence in physical character between dressed and bare excitations is evident in Figs. \ref{fig:f6}c--\ref{fig:f6}f as top and bottom panels bear little resemblance for $\lambda\gg1$. Interestingly, the ground state overlap with the dressed Mott-insulator, dressed superfluid and photonic superfluid all display regions of near-unity as $J/\hbar g$ is tuned, indicating a much more complicated phase transition in comparison to the dispersive case. Referring again to Fig. \ref{fig:f6}a, regions where the polariton-polariton repulsion strength $U_{\textrm{eff}}$ dominates the effective tunneling strength $J_{\textrm{eff}}^{(2)}$ coincide with a polaritonic Mott-insulating ground state $\ket{\widetilde{\Psi}_{\textrm{MI}}}$, as expected. In the far-opposite regime, where the effective tunneling dominates, it is evident that the dressed Mott-insulating and superfluid states fail to accurately capture the character of the ground state. Instead, it is the photonic superfluid state $\ket{\Psi_{\textrm{SF}}}$ which characterizes the ground state in the regime $J/\hbar g\gg 1$, $\lambda\gg1$. To understand this phenomenon, it is helpful to consider the original, untransformed form of the JCH Hamiltonian in Eq. (\ref{eq:twositeJCH}) where the cross-site tunneling appears in terms of purely photonic operators. For $J\gg \hbar g$, the on-site light-matter interactions contribute only perturbatively and may be neglected at first approximation. In this limit, then, the dressed operators no longer describe the fundamental excitations of the system and purely photonic character underlies the resulting superfluid-like ground state.

Remarkably, Fig. \ref{fig:f6}e indicates that a third phase, consistent with a polaritonic superfluid, appears between the regions coinciding with a polaritonic Mott-insulator and photonic superfluid for $\lambda\gg 1$. The existence of such a phase in the JCH model has been both theoretically \cite{PhysRevA.77.033801} and experimentally \cite{PhysRevLett.111.160501} examined in the literature, and may be explained as follows: as the ratio between the photonic hopping strength and light-matter coupling rate is tuned from its far limit $J/\hbar g\ll 1$ (leading to localized polaritonic excitations) to its counterpart $J/\hbar g\gg 1$ (resulting in delocalized photonic excitations), the system passes through an intermediate region $J/\hbar g\sim 1$ where $J$ is large enough such that the cross-site cavity-TLS couplings $M_2$ and $M_3$ become appreciable, yet not so large that the photonic hopping completely dominates light-matter interactions and the atomic degrees of freedom are eliminated. The result is a two particle ground state which assumes a near-unity overlap with the polaritonic superfluid state, reaching $|\braket{\widetilde{\Psi}_{\textrm{SF}}|\Psi_0}|^2\approx 0.95$ at its peak. It is interesting to note that in this parameter regime, the dynamics are not entirely restrained to the subspace $\mathcal{H}_{--}$ as was the case for dispersive coupling. Yet, the ground state is well-characterized by $\ket{\widetilde{\Psi}_{\textrm{SF}}}$ which is composed of the three individual states $\ket{2,0,-,-}$, $\ket{0,2,-,-}$, and $\ket{1,1,-,-}$, which collectively span the two excitation manifold of $\mathcal{H}_{--}$. Inspection of the excited states illustrates that this is not the case in general, indicating that quantum interference between the inter-branch transitions likely plays an important role in the system dynamics near the ground state energy.

We conclude our analysis by making a few remarks on additional phenomena of the JCH model not explored here. The preceding calculations are restricted to the case $\Delta>0$ which, as discussed in Section \ref{subsec:manybodyrep}, corresponds to repulsive on-site boson-boson interactions. Not included in the present analysis is the $\Delta<0$ limit of the two-site JCH, where attractive on-site boson-boson interactions are realized and, consequently, multiple photon (or polariton) bound states may be formed. We defer discussion of these effects to existing literature on this subject (see, for example, Refs.~\onlinecite{PhysRevA.83.055802} and ~\onlinecite{zhu2013scattering} for theoretical analyses pertaining to JCH systems and Refs.~\onlinecite{liang2018observation} and ~\onlinecite{firstenberg2013attractive} for related studies in atomic Rydberg platforms), and leave an in-depth analysis through the lens of the bosonic many-body form of the JCH model presented here as an interesting potential future avenue. Separately, dispersive coupling offers additional possibilities not explicitly discussed here, such as the realization of $XY$ spin models by either (i) mapping polariton operators onto psuedo spin operators in the Mott regime \cite{Angelakis2007, koch2009superfluid} or, alternatively, (ii) explicit separation of photonic and atomic degrees of freedom and realization of photon mediated spin-spin like interactions between the weakly dressed atoms \cite{Aron2016,Zhu_2013}. Here, the former approach is equivalent to consideration of Eq. (\ref{eq:JCH_dispersive}) in the hardcore limit $U_{\textrm{eff}}^\pm\rightarrow\infty$, while the latter involves inclusion of terms proportional to $\widetilde{\sigma}^+_1\widetilde{\sigma}^-_2 + \widetilde{\sigma}^-_1\widetilde{\sigma}^+_2$ which appear at second order in $\lambda$ in Eq. (\ref{eq:JCH_dispersive}) and play a particularly important role in the low energy dynamics of the atomic dispersive limit ($|\lambda|\ll1$ with $\Delta<0$). Thus the results presented here not only provide a direct route for comparison between the two-site JCH and Bose-Hubbard models, but also demonstrate a more general utility as a potential aid for theoretical discovery and experimental realization of other quantum Hamiltonians of interest for analog quantum simulation using cavity and circuit QED platforms.


\section{Conclusion}\label{sec:Concl}

Systems of interacting photons are among the most promising experimental platforms for studying quantum many-body phenomena. As photons do not naturally interact with each other, however,  a nonlinear element, such as a quantum emitter or superconducting qubit, is required to mediate effective photon-photon interactions. Here, we have presented a comprehensive theoretical study of the effective many-body interactions underlying the Jaynes-Cummings model, the prototypical description of light-matter coupling in cavity and circuit QED systems. This was achieved through techniques of unitary transformation, ultimately resulting in a reexpression of the Jaynes-Cummings Hamiltonian in terms of dressed bosonic and psuedo-spin operators. Upon non-perturbative expansion of its diagonal form, we have shown that the resulting dressed operator representation of the Jaynes-Cummings Hamiltonian includes an infinite sum of bosonic $k$-body interactions partitioned into two distinct branches. We have demonstrated that this many-body representation facilitates a close inspection of the parameter-dependent analogy between the Jaynes-Cummings Hamiltonian and the on-site portion of the Bose-Hubbard model. While prior studies have qualitatively compared the two -- even going so far as to define an effective Hubbard-like interaction strength $U_{\textrm{eff}}$ for the Jaynes-Cummings Hamiltonian \cite{hartmann2016quantum, koch2009superfluid, Noh2016} -- our approach is unique in that the resulting many-body form is exact for both resonant and dispersive regimes for an arbitrary number of excitations. Furthermore, our results provide a novel interpretation of the breakdown in this analogy for resonant coupling, occurring due to the emergent role of higher effective $k$-body interactions which suppress the influence of the two-body terms. These findings thus not only serve as a unique lens for comparison with the onsite interactions of the Bose-Hubbard model, but also provide a theoretical avenue for explicit study of large effective $k$-body interactions facilitated by the Jaynes-Cummings interaction for potential realization of exotic quantum behavior not realizable in conventional quantum systems \cite{Buechler2007,naidon2017efimov}.

In addition, we have extended our analysis to the two-site Jaynes-Cummings-Hubbard (JCH) model and have demonstrated that, in the dispersive coupling regime, unitary transformation to the dressed operator representation allows for a near exact realization of the two-site Bose-Hubbard model, complete with explicit, analytic forms for all effective parameters. To better understand the resonant coupling case, we then restricted to a total of two excitations or fewer, derived an explicit form for the dressed state representation of the two-site JCH, and identified the block of matrix elements which map to Bose-Hubbard-like dynamics, deriving effective two excitation hopping ($J_{\textrm{eff}}^{(2)}$) and effective two-body interaction ($U_{\textrm{eff}}$) strengths in the process. Drawing upon this theoretical foundation, we have illustrated that, for resonant coupling, the turn on of inter-branch transitions induced by cross-site dressed light-matter couplings is ultimately the downfall of analogy with the two-site Bose-Hubbard model. We then concluded with an analysis of the quantum phases of the two-site JCH model for $n=2$ excitations, illustrating the possibility for either a photonic (dispersive coupling) or polaritonic (resonant coupling) Mott-insulating state for $J_{\textrm{eff}}^{(2)}/U_{\textrm{eff}}\ll 1$, while $J_{\textrm{eff}}^{(2)}/U_{\textrm{eff}}\gg 1$ uniformly leads to a photonic superfluid state. Finally, we identified the possibility for a third quantum phase near $J_{\textrm{eff}}^{(2)}/U_{\textrm{eff}}\sim 1$ for resonant coupling, corresponding to a polaritonic superfluid-like state. While these four unique quantum phases have been identified in the literature previously \cite{greentree2006quantum, hartmann2006strongly,PhysRevA.77.033801, Noh2016}, the dressed operator picture developed here provides an explicit analytic mapping between the parameters of the JCH model and those of the effective many-body representation through which its quantum phases are easily understood, resulting in a clear, all-encompassing exposition of the various parameter regimes and their association with the quantum phases of the JCH model. The present work thus demonstrates the general utility of the dressed many-body description of the Jaynes-Cummings model and its extensions to a lattice, opening avenues for further exploration of quantum many-body phenomena realizable in coupled light-matter systems.

\begin{acknowledgments}
We  thank  Norman Yao, Marcus Bintz, and Arka Majumdar for  helpful  discussions on connecting the Jaynes-Cummings-Hubbard and Bose-Hubbard models with application to many-body quantum simulation using coupled nanophotonic cavities. This research was supported by the National Science Foundation under Grant Nos. CHE-1954393 (K.C.S., D.J.M.) and QII-TAQS-1936100 (A.B., D.J.M.).
\end{acknowledgments}

\appendix
\section{\label{sec:App1} Derivation of the many-body coefficients $C_k^\pm$}
The purpose of this appendix is to expand upon the steps taken in arriving at Eqs. (\ref{eq:Hfinal}--\ref{eq:mbcoefficients}). As mentioned in the main text, a crucial step involves Taylor expanding $f(n)$ not about small $\lambda$ as is typical for studies in the dispersive regime \cite{boissonneault2009dispersive,blais2020circuit}, but about $n=n_0$ where $n_0$ is an undetermined constant chosen to be sufficiently large such that the convergence condition $n_0>(n-1/4\lambda^2)/2$ is satisfied. Carrying out this expansion leads to
\begin{equation}
    \begin{split}
        f(n) &= \sum_{r=0}^{\infty}\binom{\frac{1}{2}}{r}(2\lambda)^{2r} f(n_0)^{1-2r}(n-n_0)^r \\
        &=\sum_{r=0}^\infty\sum_{m=0}^r\binom{\frac{1}{2}}{r}\binom{r}{m}(2\lambda)^{2r}f(n_0)^{1-2r}(-n_0)^{r-m}n^m,
    \end{split}
\end{equation}
where the binomial theorem was used in going from the first to second line. Reexpressing in terms of operators using Eq. (\ref{eq:fNsig_0}),
\begin{equation}
    \begin{split}
        f(N)\widetilde{\sigma}^z &=\sum_{r=0}^\infty\sum_{m=0}^r\binom{\frac{1}{2}}{r}\binom{r}{m}(2\lambda)^{2r}f(n_0)^{1-2r}(-n_0)^{r-m} \\
        &\times\left[(\widetilde{a}\,\widetilde{a}^\dagger)^k\widetilde{\sigma}_+\widetilde{\sigma}_- + (\widetilde{a}^\dagger\widetilde{a})^m\widetilde{\sigma}_-\widetilde{\sigma}_+\right],
    \end{split}
    \label{appeq:fNsig_0}
\end{equation}
where the commutator $[\widetilde{a},\widetilde{a}^\dagger]=1$ has been used in rewriting the projected number operator $N\sigma_+\sigma_-=\widetilde{a}^\dagger\widetilde{a} + 1$ as $\widetilde{a}\,\widetilde{a}^\dagger$. The above relation can be further rewritten using the identity \cite{blasiak2007combinatorics}
\begin{equation}
    (\widetilde{a}^\dagger \widetilde{a})^m = \sum_{k=0}^m\bracenom{m}{k}(\widetilde{a}^\dagger)^k(\widetilde{a})^k,
    \label{eq:stirlingID}
\end{equation}
where the coefficients $\bracenom{m}{k}$ are Stirling numbers of the second kind, related to the binomial coefficients via
\begin{equation}
    \bracenom{m}{k}=\frac{1}{k!}\sum_{p=0}^k\binom{k}{p}(-1)^{p-k}p^m
    \label{appeq:stirlingdef}
\end{equation}
Similarly, through combination of Eq. (\ref{eq:stirlingID}) and the binomial theorem, the following identity may be derived:
\begin{equation}
    (\widetilde{a} \,\widetilde{a}^\dagger)^m = \sum_{k=0}^m\bracenom{m+1}{k+1}(\widetilde{a}^\dagger)^k(\widetilde{a})^k.
\end{equation}
Then Eq. (\ref{appeq:fNsig_0}), using the above identities, may be written in the form given by Eq. (\ref{eq:fNsig_1}), restated here for clarity: 
\begin{equation}
     \begin{split}
        f(N)\widetilde{\sigma}^z = &\sum_{r=0}\sum_{m=0}^r \binom{\frac{1}{2}}{r}\binom{r}{m}(2\lambda)^{2r} f(n_0)^{1-2r}(-n_0)^{r-m} \\
        \times&\sum_{k=0}^m(\widetilde{a}^\dagger)^k(\widetilde{a})^k \left[\bracenom{m+1}{k+1}\widetilde{\sigma}^+\widetilde{\sigma}^- - \bracenom{m}{k}\widetilde{\sigma}^-\widetilde{\sigma}^+\right].
     \end{split}
     \label{appeq:fNsig_1}
\end{equation}
As currently written, the above expression is nearly in the desired form, containing terms proportional to the normally-ordered product $(\widetilde{a}^\dagger)^k(\widetilde{a})^k$ describing effective $k$-body bosonic interactions. In order to write a Hamiltonian as a sum over these interactions, the three nested sums of Eq. ({\ref{appeq:fNsig_1}}) must be reordered such that all $k$-body terms can be factored. Noting that the indices obey $0\leq k \leq m \leq r \leq \infty$, the ordering of the nested sums may be reversed by rewriting the upper and lower bounds, leading to Eq. (\ref{eq:Hfinal}) of the main text:
\begin{equation}
    H = \hbar\omega_c\left(N-\frac{1}{2}\right) + \sum_{k=0}^{\infty}\frac{1}{k!}\left[ C^+_k\widetilde{\sigma}_+\widetilde{\sigma}_- + C^-_k\widetilde{\sigma}_-\widetilde{\sigma}_+\right](\widetilde{a}^\dagger)^k(\widetilde{a})^k,
\end{equation}
where
\begin{equation}
    \begin{split}
        \frac{C_k^-}{k!} &= -\frac{\hbar}{2}\Delta \sum_{m=k}^\infty \bracenom{m}{k}(-n_0)^{-m}f(n_0)\sum_{r=m}^\infty\binom{\frac{1}{2}}{r}\binom{r}{m}\beta^r \\ 
        \frac{C_k^+}{k!} &= \frac{\hbar}{2}\Delta \sum_{m=k}^\infty \bracenom{m+1}{k+1}(-n_0)^{-m}f(n_0)\sum_{r=m}^\infty\binom{\frac{1}{2}}{r}\binom{r}{m}\beta^r,
    \end{split}
\end{equation}
and $\beta = -4\lambda^2n_0/f(n_0)^2$.
The above expressions may be simplified through explicit evaluation of the sum over $m$ using properties of the generalized binomial coefficients. In particular, it can be shown that
\begin{equation}
    \begin{split}
        \sum_{r=m}^\infty\binom{\frac{1}{2}}{r}\binom{r}{m}\beta^r &= \binom{\frac{1}{2}}{r}\beta^m(1+\beta)^{\frac{1}{2}-m} \\ 
        &= \binom{\frac{1}{2}}{r}(4\lambda^2)^m(-n_0)^m/f(n_0).
    \end{split}
    \label{appeq:A9}
\end{equation}
Focusing on $C_k^-$, evaluating the sum over $m$ gives
\begin{equation}
    \begin{split}
        \frac{C_k^-}{k!} &= -\frac{\hbar }{2}\Delta\sum_{m=k}^\infty\bracenom{m}{k}\binom{\frac{1}{2}}{m}(4\lambda^2)^m \\
        &= -\frac{\hbar}{2k!}\Delta\sum_{p=0}^k\binom{k}{p}(-1)^{p-k}\sum_{m=0}^\infty \binom{\frac{1}{2}}{m}(4\lambda^2p)^m\\
    \end{split}
\end{equation}
where the identity in Eq. (\ref{appeq:stirlingdef}) has been applied and the two sums reordered. Evaluating the rightmost sum (and ignoring issues of convergence as the double sum, taken together, must be convergent) yields the desired result
\begin{equation}
    C_k^- = -\frac{\hbar}{2}\Delta\sum_{p=0}^k\binom{k}{p}(-1)^{p+k}\sqrt{1+4\lambda^2p},
\end{equation}
which is identical to the form of $C_k^-$ Eq. (\ref{eq:mbcoefficients}). The derivation of $C_k^+$ follows in an analogous fashion and is therefore not made explicit here. 

Finally, we verify the form of $C_k^-$ through explicit action of the sum over all $k$-body terms on the a generic basis state $|n,-\rangle$. The methods here may again be trivially extended to verify $C_k^+$ through action on the positive branch $|n,+\rangle$. Using the properties of bosonic creation and annihilation operators,
\begin{equation}
    \begin{split}
        H_{\textrm{MB}}|n,-\rangle&=\sum_{k=0}^\infty\frac{1}{k!}C_k^{-}(\widetilde{a}^\dagger)^k(\widetilde{a})^k|n,-\rangle \\
        &=\sum_{k=0}^n\binom{n}{k}C_k^{-}|n,-\rangle.
    \end{split}
\end{equation}
Substituting the definition for $C_k^-$ and reordering the two resulting sums, again taking care to change the bounds as needed, yields,
\begin{equation}
    \begin{split}
        H_{\textrm{MB}}|n,-\rangle=-\frac{\hbar}{2}\Delta&\sum_{p=0}^{n}(-1)^p\sqrt{1+4\lambda^2p}\\
        \times&\sum_{k=p}^n\binom{n}{k}\binom{k}{p}(-1)^{k}|n,-\rangle.
    \end{split}
\end{equation}
Applying the identity
\begin{equation}
    \sum_{k=p}^n\binom{n}{k}\binom{k}{p}(-1)^k = (-1)^n \delta_{np},
\end{equation}
the above relation becomes
\begin{equation}
    H_{\textrm{MB}}|n,-\rangle=-\frac{\hbar}{2}\Delta\sqrt{1+4\lambda^2n}|n,-\rangle,
\end{equation}
thus verifying that the form of the Jaynes-Cummings Hamiltonian given in Eq. (\ref{eq:Hfinal}) returns the known eigenvalues for the negative branch states $|n,-\rangle$.

\section{\label{sec:App2} Explicit forms for $M_i$ and $K_i$}
The following lists the explicit analytic forms for the coefficients $M_i$ and $K_i$, each of which describes the amplitude of an allowed transition from the Hilbert space $\mathcal{H}_{--}$ to its complement as well as its inverse process:
\begin{equation}
    \begin{split}
        M_1 &= J\cos\theta{(1)}\sin\theta(1) \\
        M_2 &= J\sin\theta{(1)}\left[\sqrt{2}\cos\theta{(1)}\cos\theta{(2)}+\sin\theta{(1)}\sin\theta{(2)}\right] \\
        M_3 &= J\cos\theta{(1)}\left[\sqrt{2}\cos\theta{(1)}\sin\theta{(2)}-\sin\theta{(1)}\cos\theta{(2)}\right] \\
        K_1 &= J\cos\theta{(1)}\left[\sqrt{2}\sin\theta{(1)}\cos\theta{(2)}-\cos\theta{(1)}\sin\theta{(2)}\right] \\
        K_2 &= J\sin\theta{(1)}\left[\sqrt{2}\sin\theta{(1)}\cos\theta{(2)}-\cos\theta{(1)}\sin\theta{(2)}\right]
    \end{split}
\end{equation}

%

\end{document}